\documentclass[aps, pra, 10pt,twocolumn,superscriptaddress]{revtex4-2}

\usepackage[english]{babel}
\usepackage[T2A]{fontenc}
\usepackage{lmodern}
\usepackage{graphicx}
\usepackage[caption=false]{subfig}
\usepackage{amsmath}
\usepackage{amsfonts}
\usepackage{amssymb}
\usepackage{amsthm}
\usepackage{mathtools}
\usepackage{dsfont}
\usepackage{microtype}
\usepackage{tempora}
\usepackage[usenames]{color}
\usepackage{colortbl}
\usepackage{xcolor}

\usepackage{hyperref}
\hypersetup{colorlinks=true, linkcolor=blue, citecolor=blue, urlcolor=blue}

\begin{document}

\newcommand{\ket}[1]{\ensuremath{|{#1}\rangle}}
\newcommand{\bra}[1]{\ensuremath{\langle{#1}|}}
\newcommand{\sca}[2]{\ensuremath{\bigl({#1}\cdot{#2}\bigr)}}
\newcommand{\avr}[1]{\ensuremath{\langle{#1}\rangle}}

\newcommand{\cnj}[1]{{#1}^{\ast}}
\newcommand{\hcnj}[1]{{#1}^{\dagger}}
\newcommand{\tcnj}[1]{{#1}^{T}}
 
\newcommand{\prt}[1]{\partial_{#1}}
\newcommand{\pdrs}[1]{\partial_{#1}}
\newcommand{\pdr}[2]{\frac{\partial #1}{\partial #2}}
\newcommand{\drf}[2]{\frac{\dd #1}{\dd #2}}
\newcommand{\vdr}[2]{\dfrac{\delta #1}{\delta #2}}

 \def\BE{\begin{equation}}
 \def\EE{\end{equation}}
 \def\BA{\begin{array}}
 \def\EA{\end{array}}
 \def\BEA{\begin{eqnarray}}
 \def\EEA{\end{eqnarray}}
 \def\nn{\nonumber}
 \def\ra{\rangle}
 \def\la{\langle}
 \def\p{{\bf p}}
 \def\q{{\bf q}}
 \def\A{{\bf A}}
 \def\x{{\bf x}}
 \def\S{{\bf S}}
 \def\O{{\bf \Omega}}
 \def\I{{\bf I}}
 \def\U{{\bf U}}
 \def\V{{\bf V}}
 \def\Z{{\bf Z}}
 \def\d{\partial}

\title{Creation and manipulation of Schr{\"o}dinger cat states based on semiclassical predictions}

\author{N.~G.~Veselkova}
\affiliation{Saint-Petersburg State Institute of Technology, 26, Moskovski ave., St. Petersburg, 190013, Russia}
\author{R.~Goncharov}
\email[Email address: ]{rkgoncharov@itmo.ru}
\affiliation{Quantum Information Laboratory, ITMO University, Kadetskaya Line, 3, St. Petersburg, 199034, Russia}
\affiliation{Leading Research Center "National Center for Quantum Internet", ITMO University, Birzhevaya Line,
  16, St. Petersburg, 199034, Russia}
\author{Alexei~D.~Kiselev}
\email[Email address: ]{alexei.d.kiselev@gmail.com}
\affiliation{Laboratory of Quantum Processes and Measurements, ITMO
  University,199034 Kadetskaya Line 3b, Saint Petersburg, Russia} 
\affiliation{Leading Research Center "National Center for Quantum Internet", ITMO University, Birzhevaya Line,
  16, St. Petersburg, 199034, Russia}
\affiliation{Quantum Information Laboratory, ITMO University, Kadetskaya Line, 3, St. Petersburg, 199034, Russia}
% \affiliation{ITMO University, Kronverkskiy, 49, St. Petersburg, 197101, Russia}

\begin{abstract}
  We consider the generation of Schr{\"o}dinger cat states using a quantum measurement-induced
  logical gate where entanglement between the input state of the target oscillator and the Fock
  state of the ancillary system produced by the quantum non-demolition entangling $\hat{C}_Z$
  operation is combined with the homodyne measurement. We utilize the semiclassical approach to
  construct both the input-output mapping of the field variables in the phase space and the wave
  function of the output state.  \textcolor{black}{This approach is found to
    predict that the state at the gate output can be represented by a minimally
    disturbed cat-like state which is a superposition of two copies of the initial state  symmetrically
    displaced by momentum variable.} For the target oscillator prepared in the coherent state, we
  show that the fidelity between the exact solution for the gate output state and the ``perfect''
  Schr{\"o}dinger cat reconstructed from the semiclassical theory can reach high values exceeding
  0.99.
\end{abstract}

\date{\today}

\maketitle

%%%%%%%%%%%%%%%%%%%%%%
\section{Introduction}
\label{sec:introduction}
%%%%%%%%%%%%%%%

Continuous-variable (CV) optical systems are a promising platform for large-scale quantum
information processing. The properties of the CV schemes with embedded non-Gaussian gates are being
actively investigated nowadays~\cite{Moore2019, Konno2021, Hanamura2024}. Along with the large-scale
Gaussian operations enabled by the cluster states~\cite{Asavanant2021, Larsen2021}, quantum
non-Gaussian gates are essential components~\cite{Bartlett2002, Niset2009} for a variety of
practical tasks in optical quantum technology and advanced quantum information processing including
quantum communication, quantum computation, quantum algorithms, and quantum control. It has been
found that non-Gaussianity~\cite{Chitambar2019, Lee2019, Baragiola2019, Chabaud2023} in the form of
non-Gaussian quantum states \cite{Ourjoumtsev2006Quantum,Lee2002,Cooper2013, Bouillard2019,
  Tiedau2019,Quesada2019, Takase2021, https://doi.org/10.48550/arxiv.2110.03191, Walschaers2021,
  Kala2022} and non-Gaussian operations \cite{Genoni2010,Wang2015,
  Arzani2017,Zhuang2017,Sabapathy2017,Zhuang2018, Konno2021} is crucial, due to the limited
capability of Gaussian states and operations, for various CV quantum information protocols required
for quantum teleportation \cite{Fuwa2014, Zhao2023, https://doi.org/10.48550/arxiv.2404.06438},
entanglement distillation \cite{Takahashi2010, Mardani2020}, error correction \cite{Hanamura2021,
  https://doi.org/10.48550/arxiv.2305.01963}, fault-tolerant universal quantum computing
\cite{Gottesman2001, Menicucci2006, Wu2020, Konno2021, https://doi.org/10.48550/arxiv.2403.11404},
loophole-free test of quantum non-locality \cite{GarcaPatrn2004, Christensen2013}, and quantum
simulations \cite{Georgescu2014, Yanagimoto2022}.

Optical Schr{\"o}dinger cat states have been considered as a substantial non-Gaussian resource for
practical employment in quantum science since the early 2000s~\cite{Ralph2003,Gilchrist2004}. They
play a considerable role in up-to-date quantum technologies \cite{Lund2008, Mirrahimi2014,
  Neergaard-Nielsen2010, Sangouard2010, Brask2010, vanLoock2008, Goncharov2022, Joo2011,
  Gilchrist2004, Munro2002, Vlastakis2013, Tan2019} and CV quantum information
processing~\cite{Jeong2001, Gilchrist2004,Ourjoumtsev2006,Vlastakis2013}, including quantum
metrology~\cite{Joo2011,Facon2016,Knott2016,Duivenvoorden2017}, quantum
teleportation~\cite{Enk2001,Lee2013}, quantum communication and quantum
repeaters~\cite{vanLoock2008,Brask2010, Sangouard2010, Goncharov2022}, and error correction schemes
for fault-tolerant quantum
computing~\cite{Ralph2003,Lund2008,Mirrahimi2014,Gaitan2018,Albert2018,Grimsmo2020,Cai2021,
  Chamberland2022}. From a basic research point of view, the Schr{\"o}dinger cat states have been of
great interest both for testing the foundations of quantum mechanics
and determining the limits of its validity
by exploring the quantum-to-classical transition.

The main challenge for most applications in quantum information technologies is to produce
Schr{\"o}dinger cat-like states whose “size”, namely the distance in phase space between the two
coherent states, is sufficiently large to enable good-quality operations~\cite{Ralph2003, Lund2008,
  Etesse2014, Etesse2015, Sychev2017}. Such large-amplitude coherent-state superpositions exhibiting
unique non-classical attributes such as sub-Planck phase-space structures~\cite{Zurek2001} and
non-Gaussian interference features~\cite{Gerry1997,Schleich2001,Qin2021} are commonly used as a base
for preparing qubits in CV quantum computing~\cite{Cochrane1999, Ralph2003}, and a resource for
quantum error-correcting codes~\cite{Gottesman2001, Vasconcelos2010, Weigand2018, Hastrup2020}.

Currently, cat-like \textcolor{black}{coherent superpositions} are
being successfully simulated in various physical
systems~\cite{PerezLeija2016, Huang2015, Omran2019, Song2019,
  Lewenstein2021}, but developing realistic schemes to produce optical
Schr{\"o}dinger cat states with a large number of photons and
controlled quantum properties remains a challenging task. To achieve
this goal, a variety of well-established conditional generation
approaches have been proposed, including schemes based on such
non-Gaussian operations as photon-number measurement and
subtraction~\cite{Dakna1997,Dong2014,
  Gerrits2010,Asavanant2017,Takahashi2008,
  Neergaard2006,Neergaard-Nielsen2010, Bashmakova2023} and other
cat-states generation
methods~\cite{Ourjoumtsev2006Quantum,Ourjoumtsev2007,Gerrits2010,Ourjoumtsev2009,Jeong2006,Etesse2015,Sychev2017},
some of which have been successfully
implemented~\cite{Ourjoumtsev2006,Ourjoumtsev2007,Takahashi2008,
  Neergaard2006,Etesse2015,Ourjoumtsev2009,Jeong2006}. In
nondeterministic schemes which create the target state only under
predetermined conditions, the auxiliary channel can be prepared in the
Fock
state~\cite{Ourjoumtsev2007,Etesse2015,Jeong2006,Etesse2014,Sychev2019},
or even in complex superpositions which arise in the iterative cat
breeding
schemes~\cite{Etesse2015,Sychev2017,Lund2004,Laghaout2013}. In
particular, a low-frequency regime of optical Schr{\"o}dinger cat
state generation using the Fock state as a resource, a beam splitter
as an entangling element, and homodyne detection was experimentally
demonstrated~\cite{Ourjoumtsev2007,Etesse2015}.

In this paper, we consider the conditional generation of a
Schr{\"o}dinger cat state from an arbitrary coherent state employing a
two-node non-Gaussian gate (``cat gate'') based on the Fock state of
the ancillary oscillator as an elementary non-Gaussian resource, the
quantum non-demolition (QND) entangling operation $\hat{C}_Z$, and the
\textcolor{black}{projective} homodyne measurement.
\textcolor{black}{In the context of the above discussed conditional-generation platforms,
this Fock-state–based $\hat{C}_Z$ gate that not only
  achieves the fidelities comparable to those reported in leading homodyne-postselected
  schemes such as~\cite{Ourjoumtsev2007} but also delivers substantially
  higher heralding rates and flexible tuning of non-Gaussian resources
  through its measurement-dependent operation.}

\textcolor{black}{The controlled-Z (CZ) gate
  $\hat{C}_Z = e^{i \hat{q}_1 \hat{q}_2}$ is a canonical two-mode
  operation and CV analog of the two-qubit CPHASE gate, recognized as
  a fundamental CV quantum gate.
Despite
the cross-Kerr nonlinear
interactions provide a conceptual pathway for generating photon-number-dependent phase
correlations between optical modes~\cite{Grangier1998,Gerry2005},
experimental implementations of the CZ gate using nonlinear media remain
  challenging
due to weak nonlinearities and noise
  in materials like crystals/fibers~\cite{Shapiro2006}.
  Alternative experimental approaches include teleportation-based realizations
  via squeezed cluster
  states~\cite{Yoshikawa2008,Ukai2011,Shiozawa2018}.
While finite squeezing constrains the fidelity in cluster-state
implementations~\cite{Ukai2011}, these approaches leverage
deterministic generation and multiplexing capabilities of optical
systems that enable large-scale entanglement for quantum
applications~\cite{Asavanant2024}. 
}  We show that a CV quantum circuit with such
measurement-induced two-node non-Gaussian element may generate a
cat-like quantum superposition and, under optimal conditions, the
output state is close to the ``perfect'' Schr{\"o}dinger cat state
which is a superposition of the two (in general, more) symmetrically
displaced undistorted copies of the input state.  In parallel with the
exact theoretical description of the gate operation, we introduce a
clear visual interpretation of the output state based on the
semiclassical mapping of the input field variables and construct the
semiclassical wave function of the undisturbed cat state closest to
the exact output state.

This study builds upon and significantly extends the work~\cite{Baeva2024}, where the generation of
Schrödinger cat states was examined for input wavefunctions localized near the origin. Here, we
generalize this approach to arbitrary coherent states, thereby broadening the applicability of the
method. In addition, we introduce a geometric semiclassical mapping that provides an intuitive
visualization of the transformation induced by the non-Gaussian gate. By establishing a quantitative
correspondence between the exact output state and its semiclassical counterpart, we demonstrate that
the fidelity remains high across a wide range of input parameters. Furthermore, we analyze the
inherent constraints of the prior method and propose refinements that enhance the predictive
accuracy of the semiclassical description. \textcolor{black}{In contrast to
Ref.~\cite{Baeva2024} which is focused on making a comparison between the
cubic phase and Fock resource states, this manuscript provides a dedicated investigation of
Fock-state-based gates, establishing their advantages in fidelity and success probability through
both exact and semiclassical frameworks. Note that the results of a complementary study
dealing with the cubic phase resource states has been recently reported in the preprint~\cite{2504.18372}.}

The scheme can produce optical Schr{\"o}dinger cat states of any desired size with high
fidelity exceeding 0.99 and
we find the conditions under which the gate generates high-quality cat-like states by
computing the fidelity between the exact output state and the superposition of two symmetrically
displaced undistorted copies of the input coherent state.
We also demonstrate the output state quantum statistics in terms of the Wigner function.

The paper is organized as follows. In Section~\ref{section1}, we provide a semiclassical description
of the creation of Schr{\"o}dinger cat states using a measurement-induced two-node non-Gaussian
logic gate. In Section~\ref{section2}, we derive the target oscillator's exact output state
following the non-Gaussian gate's action, providing an analytic expression for the output wave
function. Section~\ref{section3} focuses on generating cat-like states from a coherent
state. Section~\ref{conclusion} concludes the paper.
Technical details on the method used to evaluate the Wigner functions
are relegated to Appendix~\ref{sec:mehler}.

%%%%%%%%%%%%%%%%%%%%%%%%%%%%%%%%%%%%%%%%
\section{Semiclassical description of Schr{\"o}dinger cat state creation}
\label{section1}
%%%%%%%%%%%%%%%%%%%%%%%%%%%%%%%%%%%%%

\begin{figure}[t!]
\centering
\includegraphics[width=0.9\columnwidth]{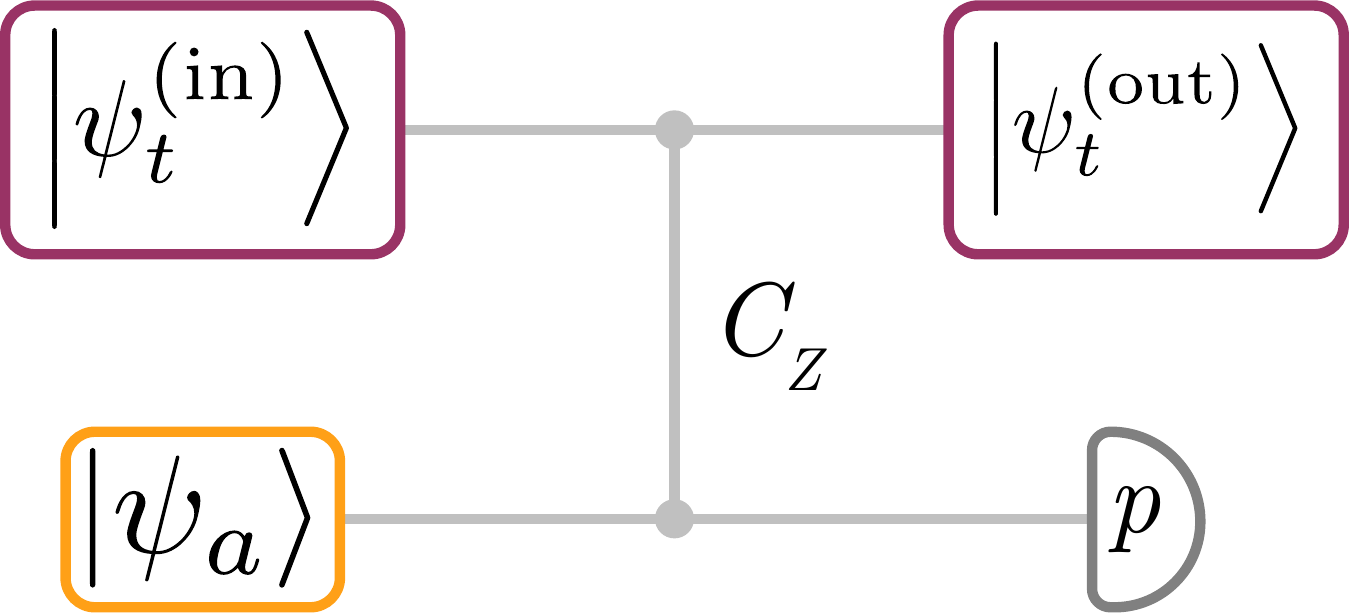}
\caption{The scheme for conditional generation of cat states employing a two-node non-Gaussian gate
  (``cat gate'') using an ancilla oscillator in the photon number (Fock) state. The states
  $|\psi^{\rm (in)}_t\rangle$ and $|\psi_a\rangle$ of the target and ancillary oscillators are sent
  to the input of the gate. After applying the entangling non-demolition $\hat{C}_Z$ operation, a homodyne
  measurement of the ancilla momentum is performed. Depending on the measurement outcome $y_m$, the
  state of the composite system collapses to a cat-like superposition state
  $|\psi^{\rm (out)}_t\rangle$.}
\label{fig1}
\end{figure}

Following to Ref.~\cite{Baeva2024},
we consider the measurement-induced two-node non-Gaussian logic gate
shown in Fig.~\ref{fig1}.
It can be seen that this gate uses 
the photon number (Fock) state of the ancilla as an elementary non-Gaussian
resource, the $\hat{C}_Z$ operation which entangles an input signal with the ancilla, and the \textcolor{black}{projective}
homodyne measurement.
%x
An essential feature of the scheme is that the ancilla measurement outcome provides multivalued
information about the target oscillator output momentum that gives rise to a cat-like output state.
This peculiarity can be easily interpreted~\cite{Sokolov2020} in terms of a clear visual
representation of the quadrature amplitudes transformations in the scheme which demonstrates that
two-component (or multi-component) Schr{\"o}dinger cat state arises when the measurement outcome is
compatible with not one but with multiple (two or more) values of the target oscillator variables. Such a
pictorial description might also be useful for the analysis of measurement-induced schemes based on
more complicated non-Gaussian resource states where
a closed-form (analytic) expression for the output state is not available.

According to Ref.~\cite{Baeva2024},
for the photon number state-based gate,
a cat-like superposition of two ``copies'' of the target state
closest to the exact output state
can be effectively evaluated using
the semiclassical approximation.
To this end,
we
assume the Heisenberg representation and 
consider
how the canonical variables, coordinate and
momentum operators, of both oscillators
are transformed under the action of the entangling operation.
Then the relations imposed on the canonical variables are treated as c-numerical and
the measurement of the ancilla momentum is also described semiclassically by substituting momentum
with its observed value giving an explicit expression for the input-output mapping
between the target oscillator variables.

\begin{figure}[t!]
\centering
\includegraphics[width=1.0\columnwidth]{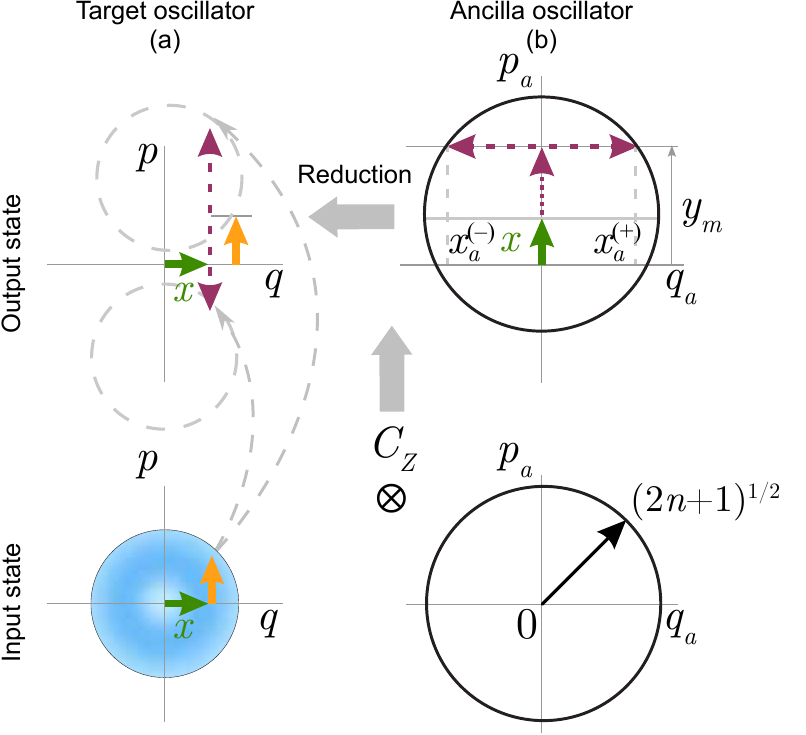}
\caption{\textcolor{black}{The scheme of the measurement-induced semiclassical mapping for the
    target oscillator ($q$, $p$; left column) and ancilla oscillator ($q_a$, $p_a$; right
    column). The target’s initial state (bottom left) is localized near the origin (circle), while
    the ancilla starts in the Fock state $|n\rangle$, represented as a circle of radius
    $\sqrt{2n+1}$ in phase space (bottom right). After the entangling operation via $\hat{C}_Z$
    gate, a homodyne measurement of the ancilla’s momentum $p_a = y_m$ (top right) splits its
    coordinate $q_a$ into two values (horizontal dashed arrows), corresponding to displacements
    $x_a^{(\pm)} = \pm \sqrt{2n+1 - (y_m - x)^2}$. This collapses the composite system into a
    two-component Schrödinger cat-like superposition state, symmetrically displaced in momentum by
    $x_a^{(\pm)}$, where $x$ is the target’s initial coordinate.}}
\label{fig2}
\end{figure}

More specifically, let us introduce the coordinate $\hat{q}$ and momentum $\hat{p}$ operators for each of the oscillators in
conventional way as $\hat{a}=(\hat{q}+i\hat{p})/\sqrt{2}$, where \textcolor{black}{$[\hat{q},\hat{p}]=i$}. Next, the two-mode entangling unitary
evolution operator \textcolor{black}{$\hat{C}_Z= \exp{(i\hat{q}\hat{q}_a)}$} is applied to the initial state of the oscillators. In the
Heisenberg picture, we have the \textcolor{black}{relations}
\textcolor{black}{
\begin{align}
  &
  \label{eq:pq-to-pq}
    \hat{q}^{\rm (out)}=\hat{q}^{\rm (in)}, \quad
    \hat{p}^{\rm (out)}= \hat{p}^{\rm (in)} +\hat{q}^{\rm (in)}_{a},
    \notag
  \\
  &
    \hat{q}^{\rm (out)}_a=\hat{q}^{\rm (in)}_a,
    \quad
    \hat{p}^{\rm (out)}_a= \hat{p}^{\rm (in)}_a +\hat{q}^{\rm (in)},
\end{align}
}
where
the index $a$ marks the ancilla variables.
In what follows,
the variables of the subsystems that enter Eq.~\eqref{eq:pq-to-pq}
will be interpreted as c-numbers.
For an ancilla initially prepared in the Fock resource state $|n\rangle$,
$n$ is the photon number,
the semiclassical amplitudes $q_a$ and $p_a$ are related as follows
\begin{align}
\label{d0}
q^{\rm (in)2}_a + p^{\rm (in)2}_a = 2n+1.
\end{align}
After applying the operation $\hat{C}_Z$,
the resource state is described by the \emph{resource curve equation}
\begin{align}
\label{d1}
q^{\rm (out)2}_a + (p^{\rm (out)}_a -q^{\rm (in)})^2 = 2n+1,
\end{align}
which is a  circle of radius $\sqrt{2n+1}$ vertically displaced by
$q^{\rm (in)}$ (see  Fig.~\ref{fig2}).

Finally, a homodyne measurement of the ancillary oscillator momentum $p^{\rm (out)}_a$
with outcome $y_m$ is described by the change $p^{\rm (out)}_a\to y_m$
giving two values of the ancilla coordinate:
$q^{\rm (out)}_{a}=x^{(\pm)}_a \equiv\pm \sqrt{2n+1-(y_m-q^{\rm (in)})^2}$.
As is shown in Fig.~\ref{fig2},
these values are the coordinates of the points of intersection
of the horizontal line $p_a^{\rm (out)}=y_m$ with the circle~\eqref{d1}.
Figure~\ref{fig2} presents a clear geometrical
description based on the semiclassical mapping that directly indicates the number,
position, and offset of these points depending on the target oscillator coordinate.

Owing the quantum correlation of the target and ancillary subsystems, 
the ambiguity of the solution for the ancilla coordinate arising from
results in the \textcolor{black}{possibility of multiple values} of the target oscillator momentum
(hereafter $x\equiv q$ denotes the target oscillator coordinate)
that enter the semiclassical mapping
\begin{align}
\label{d2}
(q^{\rm (out)},p^{\rm (out)})= (x,p^{\rm (in)}\pm \sqrt{2n+1-(y_m-x)^2}),
\end{align}
and, under additional conditions discussed below,
to the emergence of
a cat-like state which is a \textcolor{black}{coherent superposition} of two
macroscopically distinguishable coherent states. 
%

%%%%%%%%%
Note that Eq.~\eqref{d2} implies distortion of the copies of the initial state:
at any fixed $y_m$ the region of phase space where $x\approx y_m$ experiences the greatest
displacement that decreases with the magnitude of the difference $|y_m -x|$
leading to deformation of the cat components.
The distortion of copies is minimal if, for a given measurement outcome $y_m$,
the resource curve at its intersection points is $x^{(\pm)}_a$ of
the horizontal line $p_a^{\rm (out)}=y_m$ is close to vertical
(Fig.~\ref{fig2}, the top row of (b) column).
In this case, measurement-induced splitting of the ancilla coordinate (indicated by dashed
horizontal arrows) is almost independent of the target oscillator coordinate $x$ (i.e., from the
resource curve's vertical shift). In general, the ideal case is when the coordinates of the
intersection points $x^{(\pm)}_a$ do not change when shifting the position $x$ of the initial point,
chosen within the support region of the target oscillator in the phase space. For the resource Fock
state, as can be seen from Fig.~\ref{fig2}, the deformation of the copies is minimal for the phase
space points belonging to the region $x\approx y_m$---for them the horizontal dashed arrows lie on
the diameter of the Fock circle.

\begin{figure*}
\centering
  \begin{tabular}{@{}c @{} c @{} c@{}}
    \includegraphics[width=0.33\textwidth,trim={1.5cm 0.5cm 1.5cm 1cm},clip]{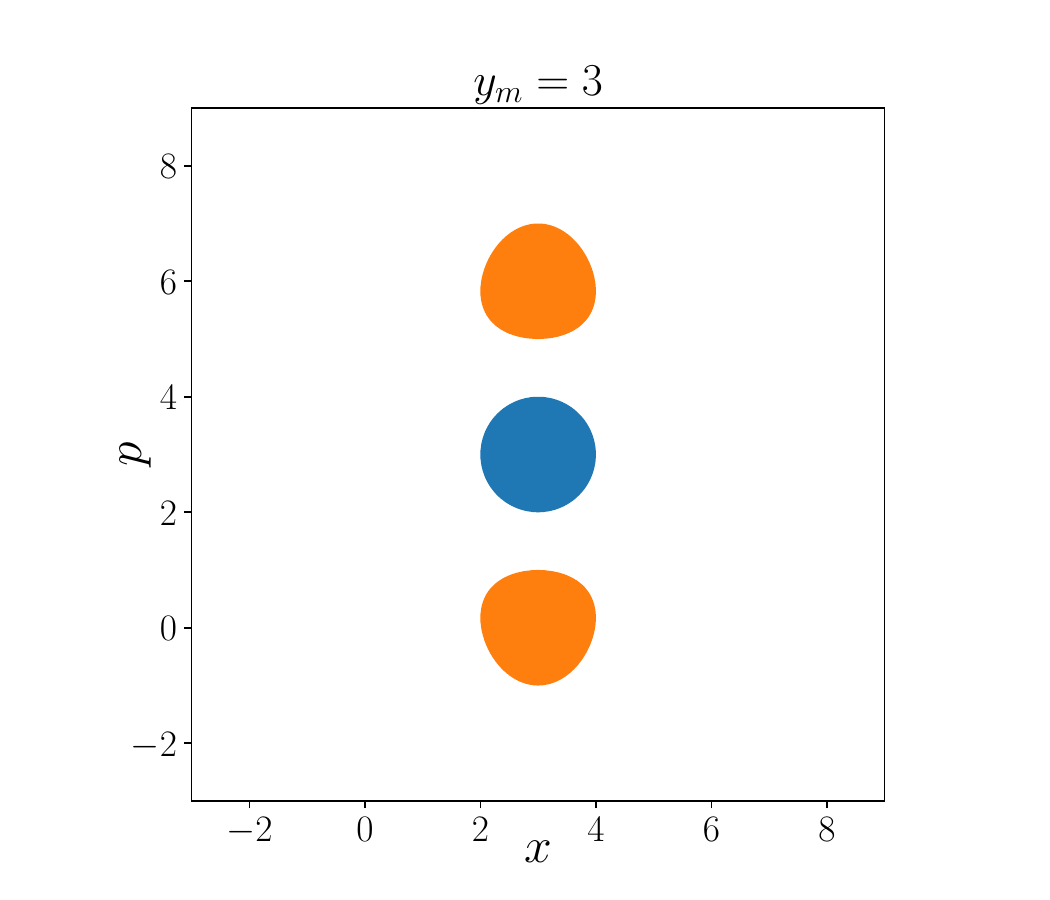} &
    \includegraphics[width=0.33\textwidth,trim={1.5cm  0.5cm 1.5cm 1cm},clip]{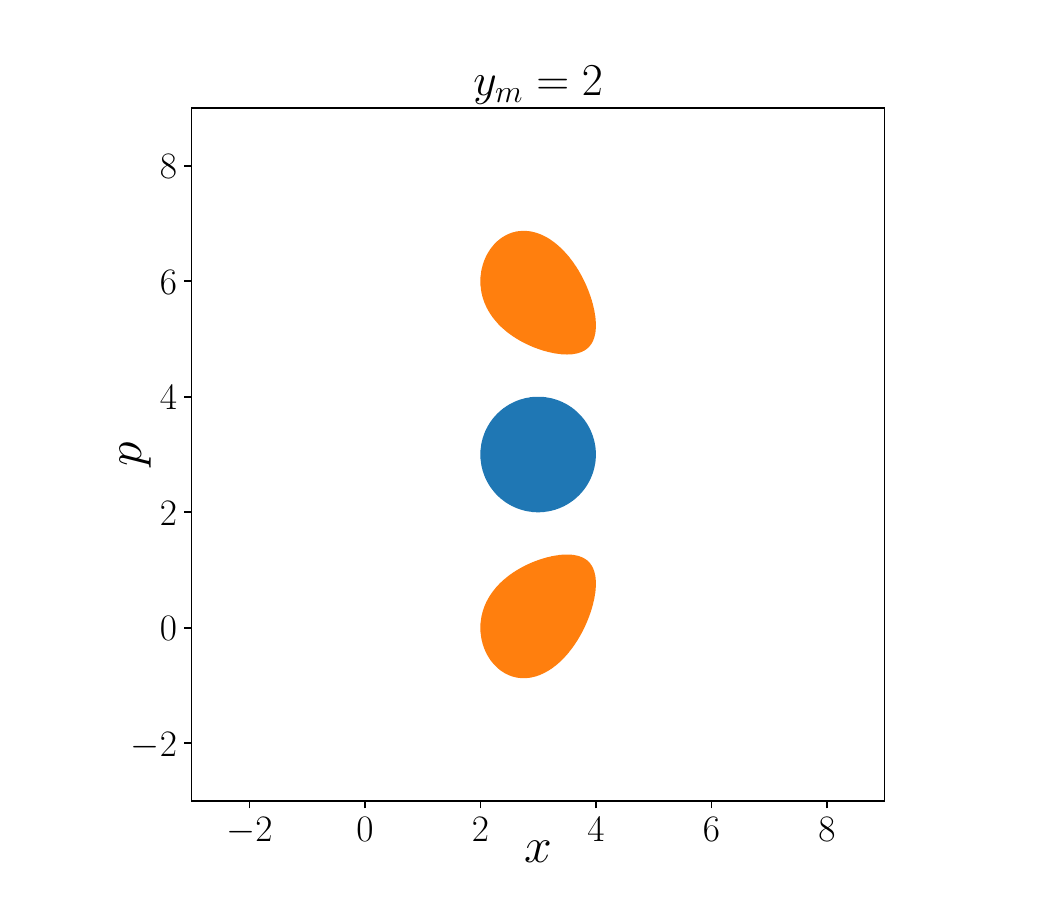} &
    \includegraphics[width=0.33\textwidth,trim={1.5cm  0.5cm 1.5cm 1cm},clip]{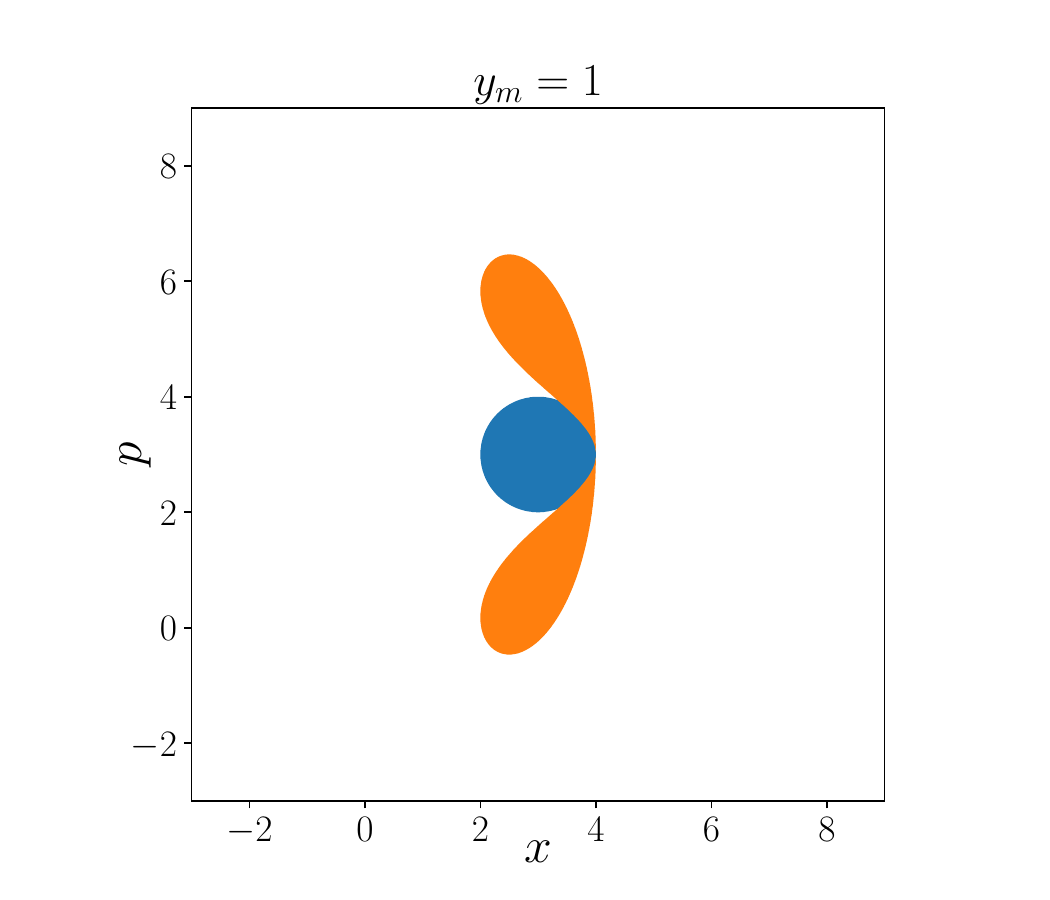}\\
    \small (a)& \small (b) & \small (c) 
  \end{tabular}
  \caption{Semiclassical mapping of quadrature amplitudes (see Eq.~\eqref{d2}) for the target oscillator prepared
    in the coherent state $\ket{\alpha}$ with the amplitude
    $\alpha=(3+3i)/\sqrt{2}$ performed by the ``cat gate'' at $n=4$ and
    (a)~$y_m =x_0= 3$; (b)~$y_m = 2$; (c)~$y_m = 1$.
    The uncertainty region of the input coherent state $(x-3)^2+(p-3)^2\le 1$ is blue colored,
    whereas its images are orange colored.
    It is demonstrated that distortions of the input
    state copies enhance with deviation of the measurement result $y_m$ from the value $x_0$
representing the coordinate of the point where the initial support region for target oscillator is
    localized.
  }
\label{fig3}
\end{figure*}

In order to illustrate these effects, let us assume that
the target oscillator is prepared in the coherent
state $\ket{\alpha}$
with the amplitude
$\alpha=(x_0+ip_0)/\sqrt{2}$ and the uncertainty region
$(x-x_0)^2+(p-p_0)^2\le 1$
indicated in Fig.~\ref{fig3} as a blue colored circle.
Figure~\ref{fig3} shows what \textcolor{black}{happens}
to the uncertainty circle under the semiclassical mapping
$(x,p)\mapsto (x,p\pm \sqrt{2n+1-(y_m-x)^2})$
(images of the circle are shown as the orange colored regions)
at different values of $y_m$.
In Fig.~\ref{fig3} the area liable to minimal deformation is:
(a)~$x\approx 3$ (the center of the uncertainty circle);
(b)~$x\approx 2$ (the left edge of the
uncertainty circle);
(c)~$x\approx 1$ (the set of points outside the uncertainty region of the
initial coherent state).

%%%%%%%%%%

The obtained semiclassical relations can be used to reconstruct the target oscillator wave
function when the ancilla oscillator in the Fock state.
Indeed, in the semiclassical treatment,
as is shown in~\cite{Baeva2024},
the input-output mapping performed by the gate under consideration
\begin{align}
\label{S101}
\psi_{\rm out}(x)= \psi_{\rm in}(x)\varphi_{\rm scl}(x),
\end{align}
can be described by multiplying the input wave function of the target oscillator $\psi_{\rm in}(x)$ by the factor
\begin{gather} 
\varphi_{\rm scl}(x)\sim \sqrt{P^{(+)}(x, y_m)}\exp{[i\int dx\,\delta p(x)]}\nn\\
+ \sqrt{P^{(-)}(x, y_m)}\exp{[-i\int dx\,\delta p(x)]},
\label{S1000}
\end{gather}
which is the sum of the added factors coming from two intersection points of the resource curve
with the horizontal line $p^{\rm (out)}_a= y_m$. It depends only on the state of the ancilla, the
target oscillator coordinate $x$, and the ancilla momentum measurement outcome $y_m$.
Formula~\eqref{S101} reflects universality of the gate implying that
its action is independent of  the target oscillator state at the input of the scheme.

From Eq.~\eqref{d2}, we have
\begin{align}
  &
\label{S111}
\delta p(x)\equiv p^{\rm (out)}- p^{\rm (in)}= \pm\sqrt{2n+1-(y_m-x)^2}
\end{align}
and, according to~\cite{Baeva2024}, the quantities $P^{(\pm)}$ given by
\begin{align}
  &
P^{(+)}(x, y_m)=P^{(-)}(x, y_m)\sim \frac{1}{|\delta p(x)|}
\label{S102}
\end{align}
can be interpreted as the classical probability of obtaining the measurement
outcome $y_m$ at the value of the ancilla coordinate in the vicinity of the overlap points
$x^{(\pm)}_a\equiv \pm \sqrt{2n+1-(y_m-x)^2}$ of the horizontal line $p^{\rm (out)}_a=y_m$ with the
resource curve $(p^{\rm (out)}_a-x)^2 +x^2_a= 2n+1$ (see Fig. \ref{fig2}(b)).

The exponential factors in Eq.~\eqref{S1000} provide the symmetrical displacement of the output
state components in phase space along the momentum axis by $\delta p(x)$.
Substituting Eq.~\eqref{S111} into the momentum produced by the gate in
the semiclassical added factor~\eqref{S1000} gives
the factor in the following explicit form:
\begin{align}
\label{f6}
\varphi_{\mathrm{\rm scl}}(n,z) \sim \frac{1}{\big(1 -  z^2\big)^{1/4}}
\big[e^{i\phi(n,z)} + (-1)^n e^{-i\phi(n,z)}\big],
\end{align}
where
\begin{align}
  &
    \phi(n,z) \equiv \frac{1}{2}(2n + 1)\left(z\sqrt{1 - z^2} +
\arcsin z\right),
    \label{f601}
  \\
  &    
z \equiv \frac{x - y_m}{\sqrt{2n + 1}}.
    \label{f600}
\end{align}

%%%%%%%%%%%%%%%%%%%%%
\section{Exact output state}
  %Schr{\"o}dinger representation} 
\label{section2}
%%%%%%%%%%%%%%%%%%%%%%%%%%%

In the Schr{\"o}dinger representation,
the action of the non-Gaussian gate schematically depicted in Fig.~\ref {fig1}
can be described based on the von Neumann reduction postulate.
This consideration yields an exact analytic expression for
the output wave function of the target oscillator
initially prepared in a quantum state that,
in the coordinate representation, is given by
\begin{align}
\label{S5}
|\psi^{\rm (in)}_t\rangle = \int dx\,\psi^{\rm (in)}(x)|x\rangle,
\end{align}
where $\psi^{\rm (in)}(x)$ is the wave function of the input state.  
The ancillary oscillator is prepared in the Fock state with $n$
photons
\begin{align}
\label{S7}
|\psi_a\rangle = \int dx_{1}\,\psi^{(n)}(x_{1})|x_{1}\rangle
\end{align}
with the wave function
\begin{align}
\label{eq:psi_n}
\psi^{(n)}(x_{1}) = \frac{1}{\pi^{1/4}\sqrt{2^n n!}}\,H_n(x_{1})e^{-x^2_{1}/2},
\end{align}
where $H_n(x_{1})$ is the Hermite polynomial.

\textcolor{black}{
A two-mode entangling QND operation $\hat{C}_Z$ applied to the composite system
state $|\psi_t^{(\mathrm{in})}\rangle\otimes|\psi_a\rangle$
leads to the state given by
}
\textcolor{black}{
\begin{align}
  \label{eq:psi_C}
  &
    \hat{C}_Z\ket{\psi_t^{\rm(in)}}\otimes\ket{\psi_a}=\int \psi^{(\mathrm{in})}\left(x\right)\psi^{(n)}\left(x_{1}\right)  
  \notag
  \\
  &
  \times e^{i x x_{1}}\left|x\right\rangle\otimes\left|x_{1}\right\rangle d x d x_{1}.
\end{align}
}
\textcolor{black}{
A subsequent projective ancilla momentum measurement 
performed on the state~\eqref{eq:psi_C}
with the outcome $p^{\rm (out)}_a=y_m$ and the corresponding momentum eigenstate
}\textcolor{black}{
\begin{align}
  \label{eq:ket-ym}
  &
    \ket{y_m}=\frac{1}{\sqrt{2\pi}}\int e^{i y_m x_1}\ket{x_1} d x_1
\end{align}
}\textcolor{black}{
results in
a reduction of the total state that can be
conveniently described in
terms of the unnormalized output state of the
target oscillator state
}\textcolor{black}{
\begin{align}
  \label{eq:tpsi-out}
  &
    \ket{\tilde{\psi}_{\rm out}}=\bra{y_m}\hat{C}_Z\ket{\psi_t^{\rm(in)}}\otimes\ket{\psi_a}
    \notag
  \\
  &
    =
    \int \psi^{(\mathrm{in})}\left(x\right)\mathcal{F}[\psi^{(n)}](x-y_m)\ket{x} d x,
\end{align}
}\textcolor{black}{
where
we have used the relation
$\avr{y_m|x_1}=e^{-iy_m x_1}/\sqrt{2\pi}$
and
$\mathcal{F}[\psi^{(n)}]$ is the Fourier transform of $\psi^{(n)}$
given by
}\textcolor{black}{
\begin{align}
  &
    \label{eq:F-trans}
    \mathcal{F}[\psi^{(n)}](p)=\frac{1}{\sqrt{2\pi}}
    \int e^{i p x_{1}} \psi^{(n)}(x_{1}) d x_1=
    i^n\psi^{(n)}(p).
\end{align}
}

So, the gate induced factor is the Fourier transform of the resource state coordinate wave
function corresponding to the value of momentum $x-y_m$ and the closed-form expression for the wave function
of the output state reads
\begin{align}
  &
\psi^{\rm (out)}(x,y_m) =\frac{1}{\sqrt{N}}{\tilde{\psi}}^{\rm (out)}(x,y_m),
    \label{S10}
  \\
  &
    \label{S010}
    {\tilde{\psi}}^{\rm (out)}(x,y_m) = \psi^{\rm (in)}(x)\
\notag
  \\
  &
  \times
\frac{i^n}{\pi^{1/4}\sqrt{2^nn!}}H_n(x-y_m)e^{-(x-y_m)^2/2},    
\end{align}
where
${\tilde{\psi}}^{\rm (out)}(x,y_m)$ is the unnormalized output wave function of the target
oscillator and $N$ is the normalization factor.
Formulas~\eqref{S10} and~\eqref{S010} define the state that will be referred to as
the \emph{exact output state}.

Note that, similar to the semiclassical formula~\eqref{S101}, the output wave function
is obtained by multiplying the input wave function by the factor representing the gate action.
This factor is solely determined by the initial state of the ancilla and
the difference between the variables $x$ and $y_m$
while remaining independent of the initial state of the target oscillator.
% The solution~\eqref{S10} can be expressed in terms of
% the Fourier transform $F$ of the Fock state wave function
% \begin{align}
%   &
% \psi^{\rm (out)}(x,y_m) \sim \psi^{\rm (in)}(x)[F\psi^{(n)}](y_m-x),
% \label{S11}
% \end{align}
% so that the gate induced factor is the Fourier transform of the resource state coordinate wave
% function corresponding to the value of momentum $y_m-x$.
In the geometric representation shown in Fig.~\ref{fig2},
the latter is consistent with the upward shift of $x$ in the circle representing
the Fock resource state on the phase plane (see top row of (b) column in Fig.~\ref{fig2}).

In the next section we demonstrate that,
when the input state is a Glauber coherent state
and the value of the measurement outcome is optimal,
the exact output state~\eqref{S10} will be close to
the Schr{\"o}dinger cat state with high fidelity
regardless of the photon number of the resource Fock state.   

%%%%%%%%%%%%%%%%%%%%%%%%%%%%
\section{Generation of cat-like states from coherent state}
\label{section3}

\begin{figure}[t!]
\centering
\includegraphics[width=1.0\columnwidth]{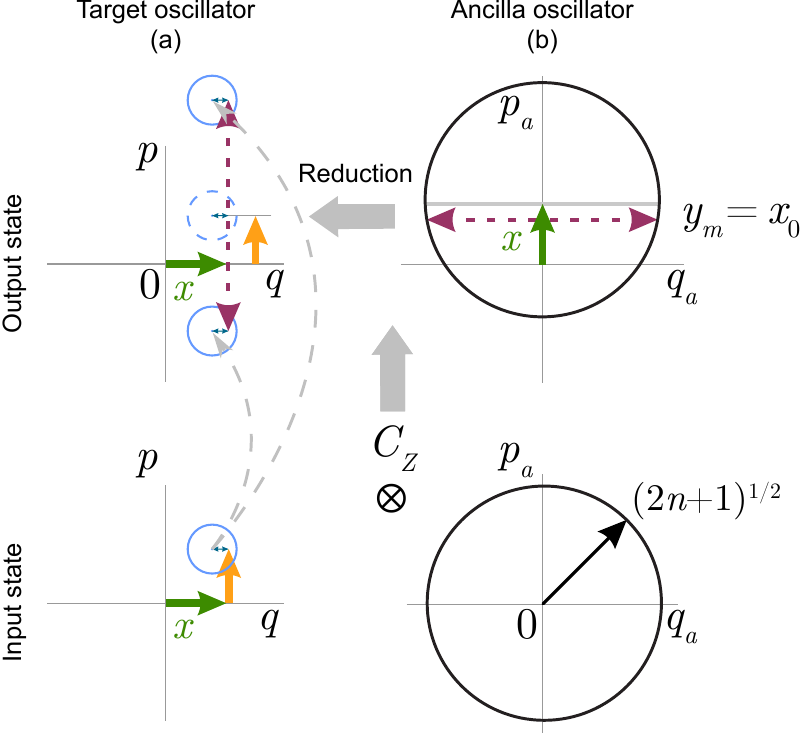}
\caption{
Schematic representation of the measurement-induced two-node ``cat gate'' operation using the Fock
state as an elementary non-Gaussian resource, the entangling $\hat{C}_Z$ operation, and the \textcolor{black}{projective}
homodyne measurement, in the form of semiclassical mapping of the quadrature amplitudes of the
target $q$, $p$ [(a) column] and ancilla $q_a$, $p_a$ [(b) column] oscillators on the phase plane
when the target oscillator is initially prepared in a coherent state $|\alpha_0\rangle$ with
amplitude $\alpha_0 = (x_0 + ip_0)/\sqrt{2}$ depicted by an uncertainty disc centered at the point
$(x_0, p_0)$ [bottom row of (a) column], and the ancilla in the Fock state with the number of
photons $n$ represented by a circle of radius $\sqrt{2n+1}$ in phase space [bottom row of (b)
column]. As a result of applying the two-mode entangling operator $\hat{C}_Z$ and carrying out a homodyne
measurement of the ancilla momentum with the outcome $y_m$, the Fock state-based gate conditionally
generates a Schr{\"o}dinger cat state—a two-component cat-like superposition state, the components
of which are symmetrically displaced in phase space along the momentum axis by $x^{(\pm)}_a = \pm
\sqrt{2n+1 - (y_m - x)^2}$, where $x$ is the initial coordinate of the target oscillator [the top
row of (a) column]. The Schr{\"o}dinger cat state arises when two values of the target oscillator
momentum correspond to the measurement result $y_m$.}  
\label{fig4}
\end{figure}

In this section,
we consider the conditional generation of
the Schr{\"o}dinger cat state via the ``cat gate'' described above
and concentrate on the special case where
the initial state of the target oscillator is  the Glauber coherent state $\ket{\alpha_0}$
with the amplitude $\alpha_0 = (x_0 + ip_0)/\sqrt{2}$,
whereas the ancillary one is still assumed to be prepared in the $n$-photon Fock state.
In the $qp$ phase space, the coherent state $\ket{\alpha_0}$ can be represented
by the uncertainty (localization) region bounded by the circle of the radius
$\delta x\sim 1/\sqrt{2}$ centered at the point $(x_0,p_0)$.
In this case,
the wave function of the input state is
\begin{align}
  &
    \label{f65}
  \psi^{\rm (in)}(x)=\avr{x|\alpha_0}=\psi_{vac}(x-x_0)\exp{\{ip_0x - ip_0x_0/2\}}
  \notag
  \\
  &
=\pi^{-1/4}\exp{\{-(x-x_0)^2/2 + ip_0x - ip_0x_0/2\}},
\end{align}
where $\psi_{vac}$ is the vacuum wave function in the coordinate representation, 
and Figure~\ref{fig4} schematically shows
the semiclassical input-output mapping of the quadrature amplitudes of the target and
auxiliary oscillators predicting an emergence of the output state in the form of a cat-like
superposition of two ``copies'' of the input target state.

\begin{figure*}[t!]
\centering
\includegraphics[width=1.0\textwidth]{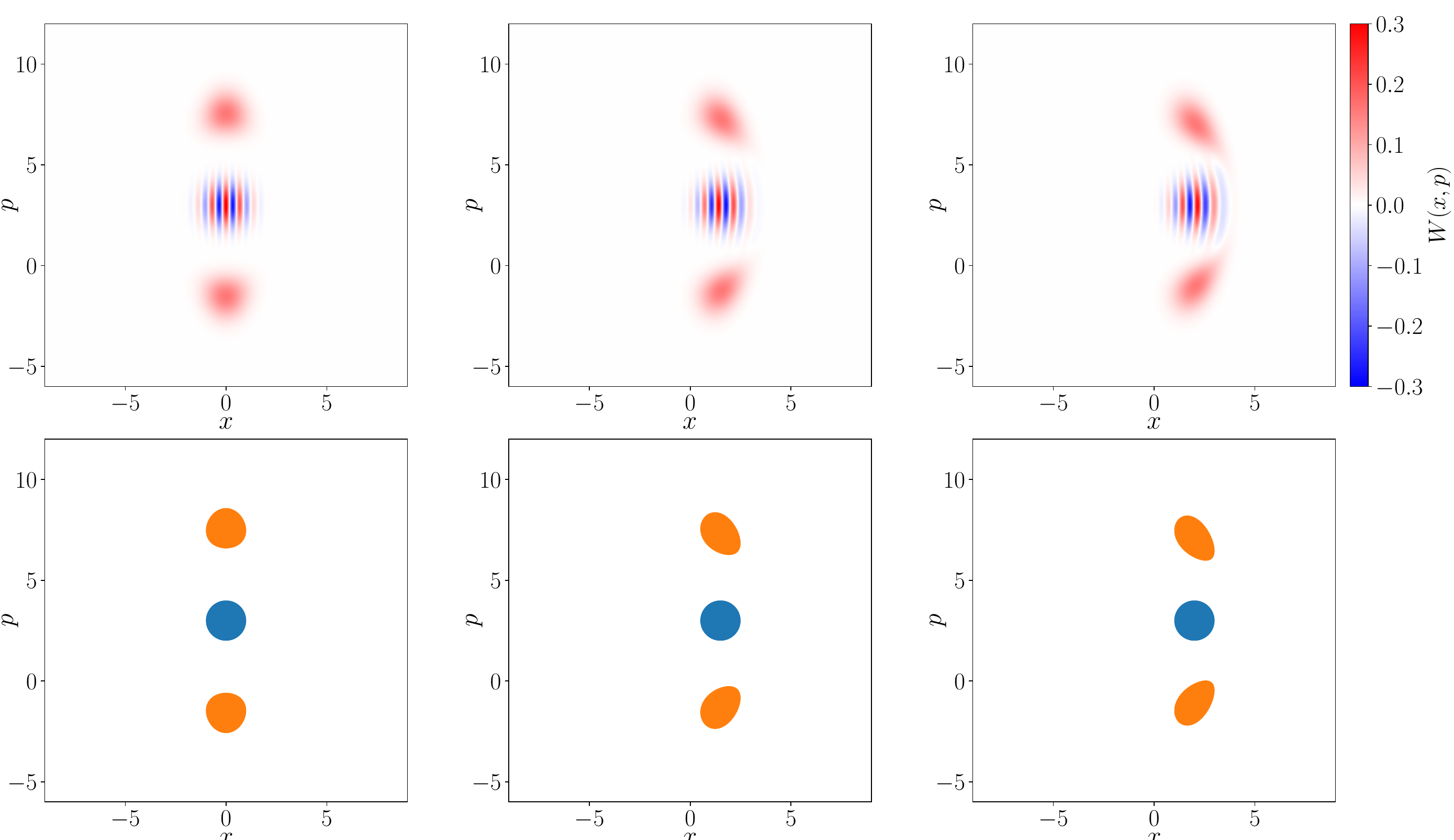}
\caption{The (top row)~Wigner function of the exact output state~\eqref{S10}  and
  (bottom row)~semiclassical
  mapping of the quadrature amplitudes of the target oscillator
  (see Eq.~\eqref{d2}) at $n=10$ and $y_m = 0$ for the input coherent state~\eqref{f65} 
  with $p_0=3$ and (left column)~$x_0=0$;
  %- the generated state is a minimum disturbed Schr{\"o}dinger cat;
(central column)~$x_0=1.5$; (right column)~$x_0=2$.
%$p_0$ everywhere chosen to be equal to $3$ does not affect the quality of the cat-like superposition generated by the gate.
The radius of the uncertainty region for the input coherent state is unity.
See also the caption of Fig.~\ref{fig3}.
}
\label{fig5}
\end{figure*}

%the generated state is a minimally disturbed Schr ¨odinger cat
In Fig.~\ref{fig5}, we compare the Wigner function of
the exact output state~\eqref{S10}
\begin{align}
  &
\label{f5}
W(x,p) =
    \frac{1}{\pi}\int dz\,\psi^{\rm ({out})*}(x+z,y_m)
    \notag
  \\
  &
    \times
    \psi^{\rm({out})}(x-z,y_m)e^{2ipz}
\end{align}
computed  for the input coherent state~\eqref{f65}
with the appropriate semiclassical mapping of
the phase space region corresponding to the Gaussian function~\eqref{f65} performed by the gate
according to the Eq.~\eqref{d2}.
Calculations were performed
at $y_m = 0$ for the photon number $n=10$
with $x_0\in\{0,1,2\}$ assuming that
the uncertainty region radius of the mapped coherent state is unity.
For calculating the Wigner functions,
we have developed the fast and efficient method 
which is based on the technique of generating functions and, thus,
avoids performing time consuming numerical integration.
Details on this method are relegated to Appendix~\ref{sec:mehler}.

Referring to Fig.~\ref{fig5},
results for the Wigner functions
and the semiclassical mapping are in good agreement. Thus, a simple geometric representation in the phase space based on the semiclassical description \textcolor{black}{reproduces} the output state with high accuracy and also \textcolor{black}{predicts} the appearance of a minimum disturbed cat-like superposition at the gate output (see the left column in Fig.~\ref{fig5}).

%
 
%%%%%%%%%%%%%%%%%
Given the input state, from Eqs.~\eqref{S101} and~\eqref{f6}, \textcolor{black}{considering that
  the input state in the phase space occupies a limited range of the coordinate near the point $x_0$
  and the weighting factor in Eq.~\eqref{f6} can be assumed constant}, we can construct the
semiclassical wave function at the gate output
\begin{align}
\label{0f7}
\psi^{\rm (out)}_{\rm scl}(x,y_m)\sim\psi^{\rm (in)}(x)\big[e^{i\phi(n,z)} + (-1)^n e^{-i\phi(n,z)}\big],
\end{align}
where $\phi(n,z)$ is the phase function  given by Eq.~\eqref{f601},
and calculate the fidelity $F_{\rm scl}$ between
the semiclassical and exact output states  
\begin{align}
\label{00f7}
F_{\rm scl}(y_m,x_0,n)= \left|\int dx\,\psi^{\rm (out)*}(x,y_m)\psi^{\rm (out)}_{\rm scl}(x,y_m)\right|^2.
\end{align}
Note that, for the input wave function~\eqref{f65}, this fidelity is independent of $p_0$.

\begin{figure}[t!]
\centering
\includegraphics[width=1.1\columnwidth]{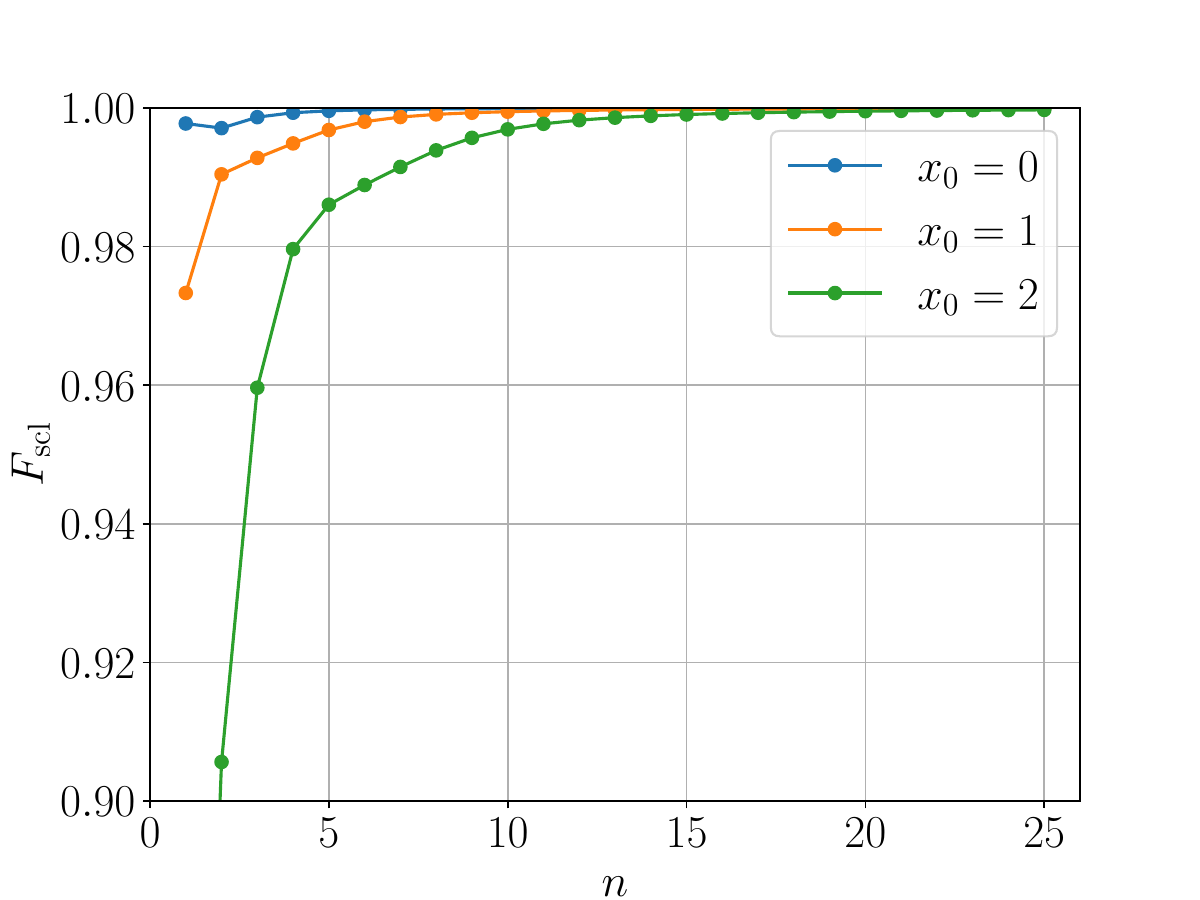}
\caption{Fidelity $F_{\rm scl}$ between the exact output state~\eqref{S10} and
  \textcolor{black}{the semiclassical wave function at the gate output}~\eqref{0f7} as a function
  of the photon number $n$ of the ancillary oscillator state at $y_m=0$ for the input coherent state
  with different values of $x_0$.}
\label{fig6}
\end{figure}

The plots of $F_{\rm scl}$ computed in relation to the photon number $n$ of the resource state at $y_m=0$
and different values canonical variable $x_0$ are shown in Fig.~\ref{fig6}.
It is seen that
the exact output state and the output state recovered from the semiclassical theory are in perfect
agreement at small $x_0$, namely, with fidelity $F_{\rm scl}> 0.9970$ at $x_0=0$ for any $n$,
$F_{\rm scl}> 0.9733$ at $x_0=1$ for any $n$, $F_{\rm scl}> 0.9056$ at $x_0=2$ for $n\ge 2$. 
In subsequent sections, 
we shall use formula~\eqref{0f7} to deduce the expression for
the semiclassical wave function of a Schr{\"o}dinger cat-like superposition of
two ``copies'' of the initial target state
representing the state closest to the exact output state~\eqref{S10}
for an arbitrary coherent input state. 

%%%%%%%%%%%%%%%%%%%%%%

\subsection{Schr{\"o}dinger cat-like state when the input wave function is localized near zero}

We begin with the case of cat state generation by the Fock state-based gate
when the input state in the phase space occupies
a limited range of the coordinate near the point $x_0 = 0$, i.e., $x\approx x_0\ll\sqrt{2n+1}$.
Under this assumption, the function $\phi(n, z)$ can be decomposed into a Taylor series in
powers of $x$, which converges at $|x|<\sqrt{2n+1}$ provided that $|y_m|\ll\sqrt{2n+1}$.
For $x\ll\sqrt{2n+1}$, we can limit ourselves to the first few terms of the
Taylor series for phase $\phi$.
So, up to the second order in the coordinate $x$,
we have 
\begin{align}
\label{f7}
\phi(n, z) \approx  \theta + p^{(+)}x +\delta p^{(+)}x^2,
\end{align}
where
\begin{align}
  &
\label{f8}
  \theta \equiv \phi\left(n, -y_m/\sqrt{2n + 1}\right), \quad p^{(+)} \equiv\sqrt{2n + 1 - y_m^2},
  \notag
  \\
  &
\delta p^{(+)}\equiv\frac{y_m}{2\sqrt{2n + 1 - y_m^2}}.
\end{align}
When the measurement result $y_m=0$ the Taylor series~\eqref{f7} can be written up to a linear term
in the coordinate $x$. This approximation yields the gate output state~\eqref{0f7} in the form 
\begin{align}
\label{f9}
\psi^{\rm (out)}_{\rm cat}(x) \sim \psi^{\rm (in)}(x)[e^{i(\theta+p^{(+)}x)}+(-1)^ne^{-i(\theta+p^{(+)}x)}],
\end{align}
which corresponds to the ``perfect''  Schr{\"o}dinger cat state, namely a superposition of two
undistorted copies of the target oscillator initial state symmetrically shifted in
the phase $q-p$ space  along the momentum axis by $\pm p^{(+)}=\pm\sqrt{2n + 1}$,
with the phase $\theta=0$.
Such semiclassical ``perfect'' cat state may be represented
in terms of a superposition of the Glauber coherent states
$|\alpha_+\rangle$ and $|\alpha_-\rangle$, 
\begin{align}
\label{f10}
|\psi^{\rm (out)}_{\rm cat}\rangle =\frac{1}{\sqrt{\cal N}}[e^{i\theta}|\alpha_+\rangle + (-1)^n e^{-i\theta}|\alpha_-\rangle],
\end{align} 
where $\alpha_{\pm} = [x_0+i(p_0\pm p^{(+)})]/\sqrt{2}$ and ${\cal N}$ is the normalization factor.

The terms of the second order and higher concerning $x$ in the Taylor series of the added factor
phase $\phi(n, z)$, indicate the dependence of the momentum transmitted by the gate on the target
oscillator coordinate $x$ and lead to a distortion of the shape of the region on the phase plane
where the Wigner function component is localized.
In the geometrical description
(see Figs.~\ref{fig2} and~\ref{fig4}),
this dependence follows from the fact that, for a given measurement
outcome $y_m$, the intersection points $x^{(\pm)}_a$ are displaced when the resource curve is shifted
along the vertical axis in changing the coordinate $x$ of the target oscillator if the resource
curve is not a vertical line at the intersection with the horizontal line $p_a=y_m$. This results in
the copies of the input state of the target oscillator undergo shear deformation of the opposite
sign due to the nearly linear dependence of the displacement on $x$ for large enough measurement
outcomes $y_m$, which is described by the quadratic term in the Taylor expand of $\phi(n, z)$ and
characterized by the measure of the linear shear deformation being $y_m/(2\sqrt{2n + 1 - y_m^2})$,
as follows from Eq.~\eqref{f8}. 
\begin{figure}[t!]
\centering
\includegraphics[width=1.1\columnwidth]{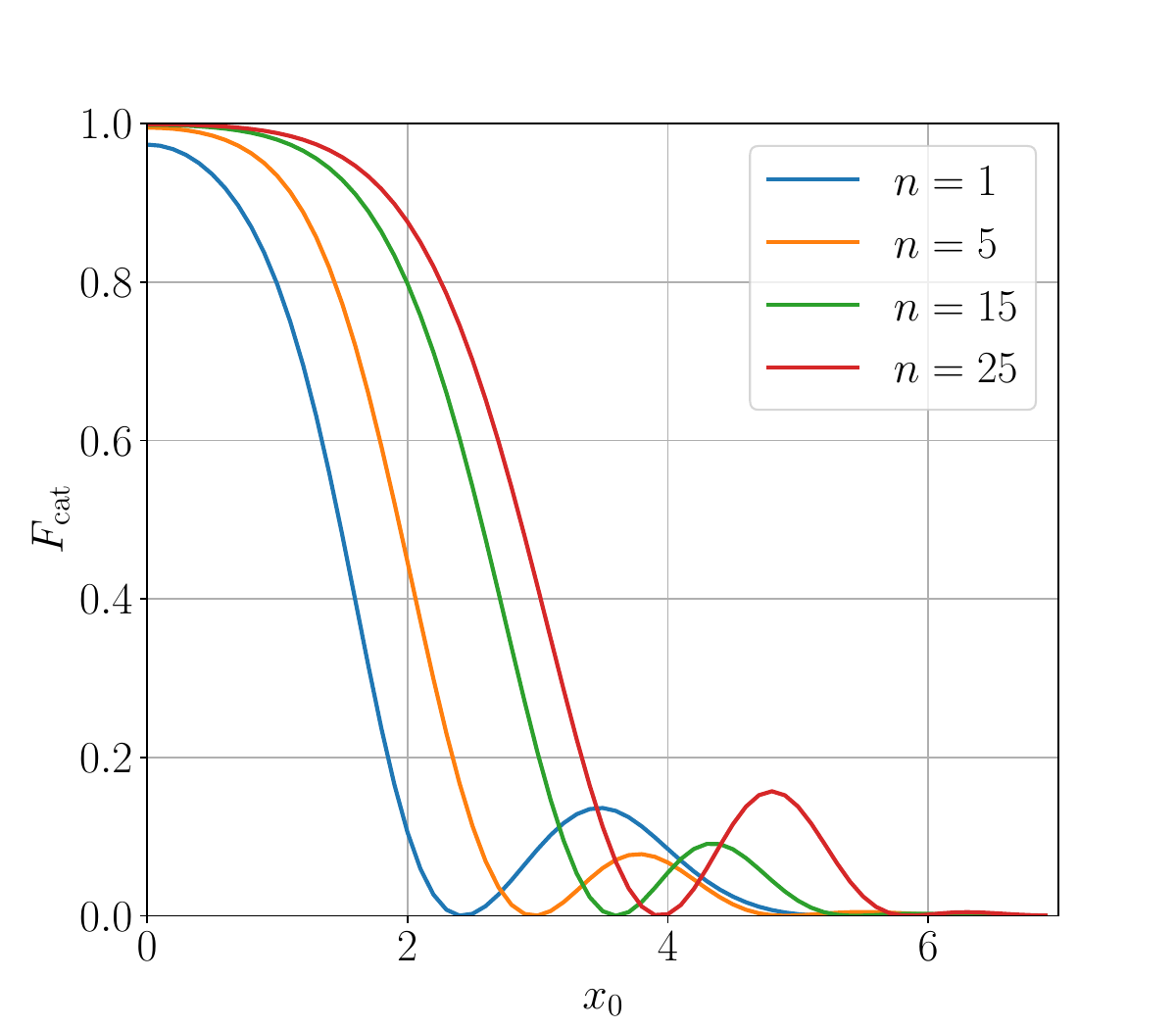}
\caption{Fidelity $F_{\rm cat}$~\eqref{D11} between the exact output state~\eqref{S10}
and the undisturbed  semiclassical cat~\eqref{f9} for the input
  coherent state~\eqref{f65} as a function of $x_0$
  at the measurement outcome $y_m=0$   and the photon number $n\in \{5,15,25\}$ of the Fock resource
  state. Fidelity $F_{\rm cat}$ is an even function of $x_0$.} 
\label{fig7}
\end{figure}

\begin{figure}[t!]
\centering
\includegraphics[width=1.1\columnwidth]{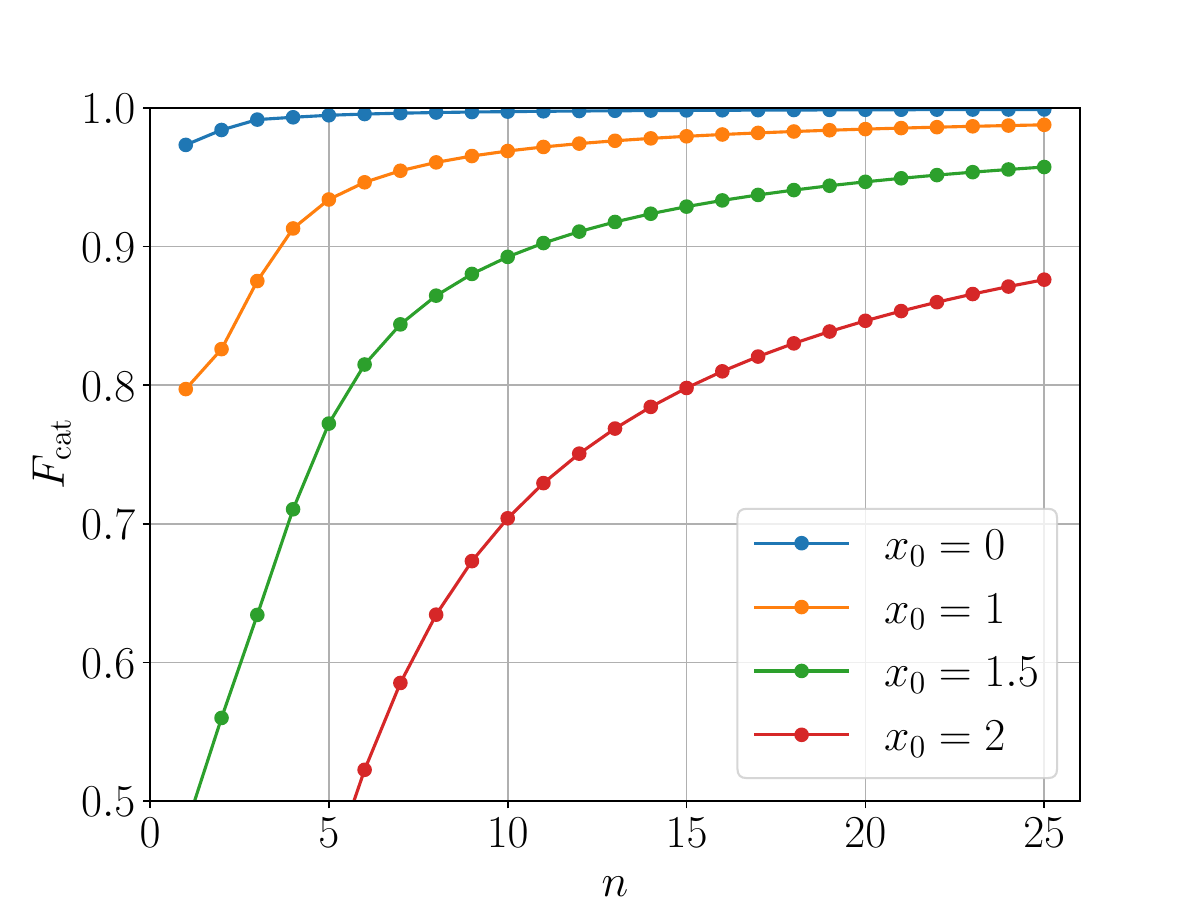}
\caption{Fidelity $F_{\rm cat}$~\eqref{D11} between the exact output state~\eqref{S10}
  and the ``perfect'' semiclassical cat~\eqref{f9}
  versus the photon number $n$ of the ancillary oscillator state
  at the measurement outcome $y_m=0$  for the input
  coherent state~\eqref{f65} with $x_0\in\{0,1,1.5,2\}$.} 
\label{fig8}
\end{figure}
In order to evaluate the ``\textcolor{black}{proximity}'' of the exact output state to the Schr{\"o}dinger cat state,
we consider the fidelity
\begin{align}
  &
\label{D11}
F_{\rm cat}(y_m,x_0,n)= \left|\int dx\,\psi^{\rm (out)*}(x,y_m)\psi^{\rm (out)}_{\rm cat}(x)\right|^2.
\end{align}
between the state~\eqref{S10} and the ``perfect'' cat state~\eqref{f9} recovered from the
semiclassical picture.

From the explicit expressions for the wave functions $\psi^{\rm (out)}(x,y_m)$ and
$\psi^{\rm (out)}_{\rm cat}(x)$, it is not difficult to conclude that $F_{\rm cat}$
does not depend on the coordinate $y_0$.
In Fig.~\ref{fig7},
$F_{\rm cat}$ is plotted against
$x_0$ at $y_m=0$ for various values of the photon number.
It is shown that the fidelity reaches \textcolor{black}{values close to unity} at $x_0=0$ for any value of $n$. So, $F_{\rm cat}$ remains close to unity only for such coherent states
where $x_0$ lies in the vicinity of the origin.
It can also be seen that the fidelity rapidly drops with $x_0$. 

The Wigner functions of the exact solution~\eqref{S10} at $y_m=0$ for the input coherent state with
$x_0\in\{0,1.5,2\}$ pictured in the top row Fig.~\ref{fig5} 
demonstrate that, for 
the resource Fock state with $n=10$,
the output state  is close to the undisturbed
Schr{\"o}dinger-cat-like superposition~\eqref{f9} at $x_0=0$
in agreement with the
curves for $F_{\rm cat}$~\eqref{D11} plotted in Figs.~\ref{fig7} and~\ref{fig8}.
Note that, for the columns shown in Fig.~\ref{fig5},
the values of the fidelity $F_{\rm cat}$
are: $0.9974$ (left column),
$0.8926$ (central column) and $0.704$ (right column).

% For the given values of the canonical variable $x_0=0,\;1.5, \;2$, the fidelity  is
% equal to
% $0.9974,\;0.8926,\;0.7040$, respectively. 

%%%%%%%%%%%%%%%%%%%%%%%%%%%%%%%%%%%%
 
\subsection{Schr{\"o}dinger cat-like state when the input wave function is localized away from zero}

From Figs.~\ref{fig7} and~\ref{fig8},
at the measurement outcome $y_m=0$,
the fidelity $F_{\rm cat}$ between the exact
solution~\eqref{S10}  and the ``perfect'' semiclassical
cat~\eqref{f9} for the input state~\eqref{f65} is close to unity
when  the $x$-quadrature $x_0$ of the input coherent state of amplitude
$\alpha_0$ is in the immediate vicinity of zero.
It is expected that the size of the vicinity is of the order of vacuum fluctuations
level $\delta x=1/\sqrt{2}$, i.e., $|x_0| \le \delta x$ (the maximum magnitude for $x_0$ can be
chosen based on the requirements for the value of $F_{\rm cat}$).

However, the fidelity
$F_{\rm cat}$ rapidly declines with $x_0$
and \textcolor{black}{the main reason} for this significant reduction is that the wave function~\eqref{f9} of the
semiclassical cat state obtained in the previous subsection
fails to give a good approximation for the output state~\eqref{S10}
provided that the magnitude of $x_0$ exceeds $\delta x$.  

In this section, we give a prescription for constructing the Schr{\"o}dinger cat state closest to the exact
output state for the general case when
the coherent state~\eqref{f65} with $x_0\ne 0$ is sent to the input of the Fock state-based gate.
In other words, the support of the input wave function
$\psi^{\rm (in)}(x)$ is now localized in the vicinity of the point $x_0$
which is well separated from the origin of the phase space.

To construct a cat-like superposition state
giving a high fidelity approximation of the exact output state,
we expand the phase~\eqref{f601} in
the added factor $\varphi_{\rm scl}$ in a Taylor series in the localization region of the Gaussian $\psi^{\rm (in)}(x)$ as was
performed in the previous subsection, i.e., now in the vicinity of the point $x=x_0$. By analogy
with~\eqref{f7}, in the region $|x-x_0|<\sqrt{2n+1}$ under the condition $|x-y_m|\ll\sqrt{2n+1}$ we
expand the phase function $\phi(n, z)$ into the Taylor series in powers of $x-x_0$.
In the sufficiently small neighborhood of $x_0$ with $|x-x_0|\ll\sqrt{2n+1}$, the expansion of the
function $\phi(n, z)$ can be truncated up to the second order terms quadratic in $x-x_0$.
So, we have
\begin{align}
\label{D3}
\varphi_{\rm scl}\sim e^{i\phi(n, z)} + (-1)^n e^{-i\phi(n, z)},
\end{align}
\begin{align}
% \label{D3}
\phi(n,z)\approx \theta_0 + p^{(+)}_0\cdot (x-x_0) +\delta p^{(+)}_0\cdot (x-x_0)^2,
\end{align}
where
\begin{align}
  &
\theta_0 =\phi(n, (x_0-y_m)/\sqrt{2n + 1}), 
    \label{D4}
  \\
  &
p^{(+)}_0 =\sqrt{2n + 1 - (x_0-y_m)^2},
    \label{D40}
  \\
  &
\delta p^{(+)}_0= -\frac{x_0-y_m}{2\sqrt{2n + 1 - (x_0-y_m)^2}}.
\label{D41}    
\end{align}

\begin{figure*}
\centering
  \begin{tabular}{c @{\qquad} c @{\qquad} c}
    \includegraphics[width=0.25806\linewidth]{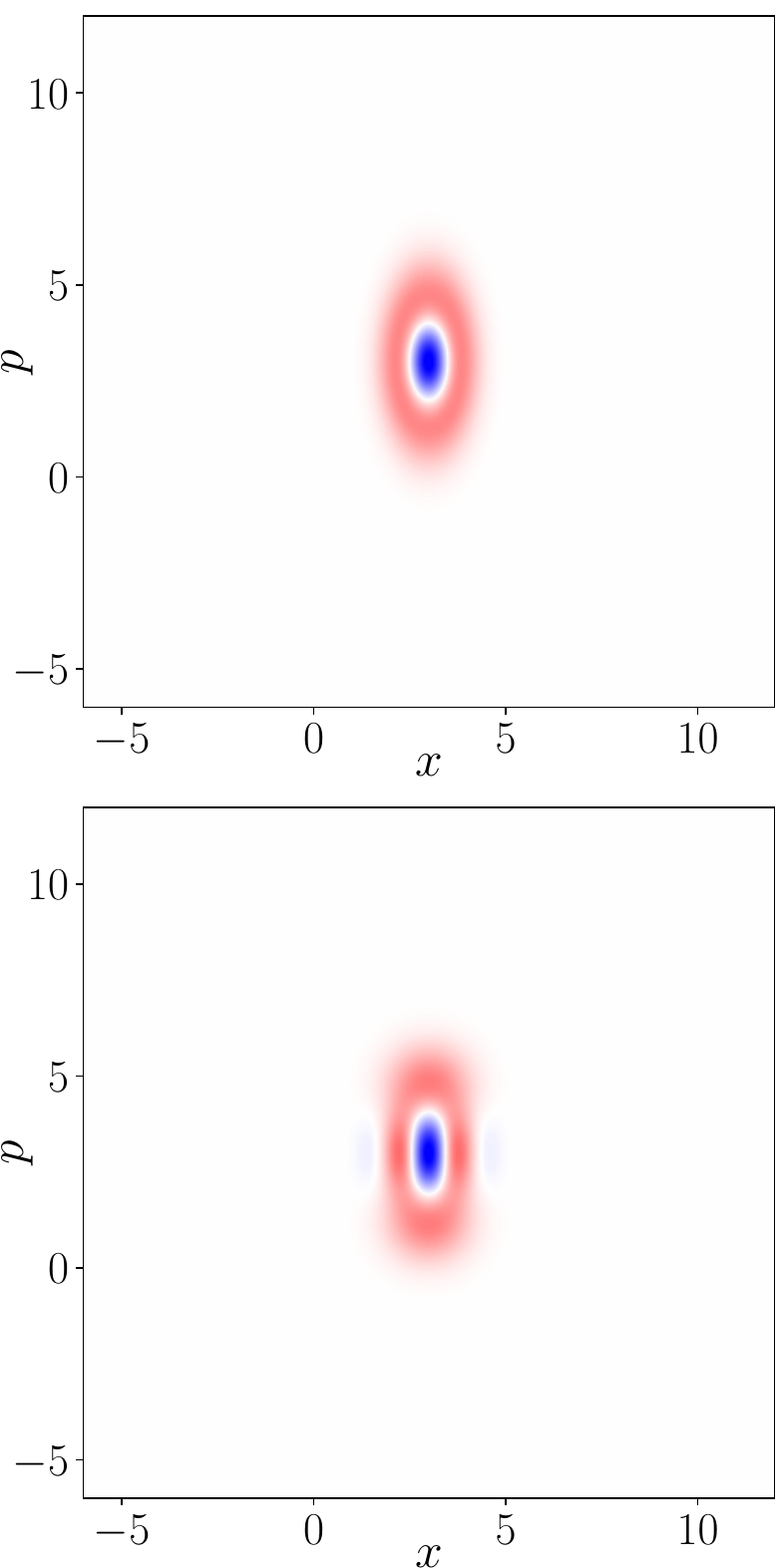} &
    \includegraphics[width=0.25806\linewidth]{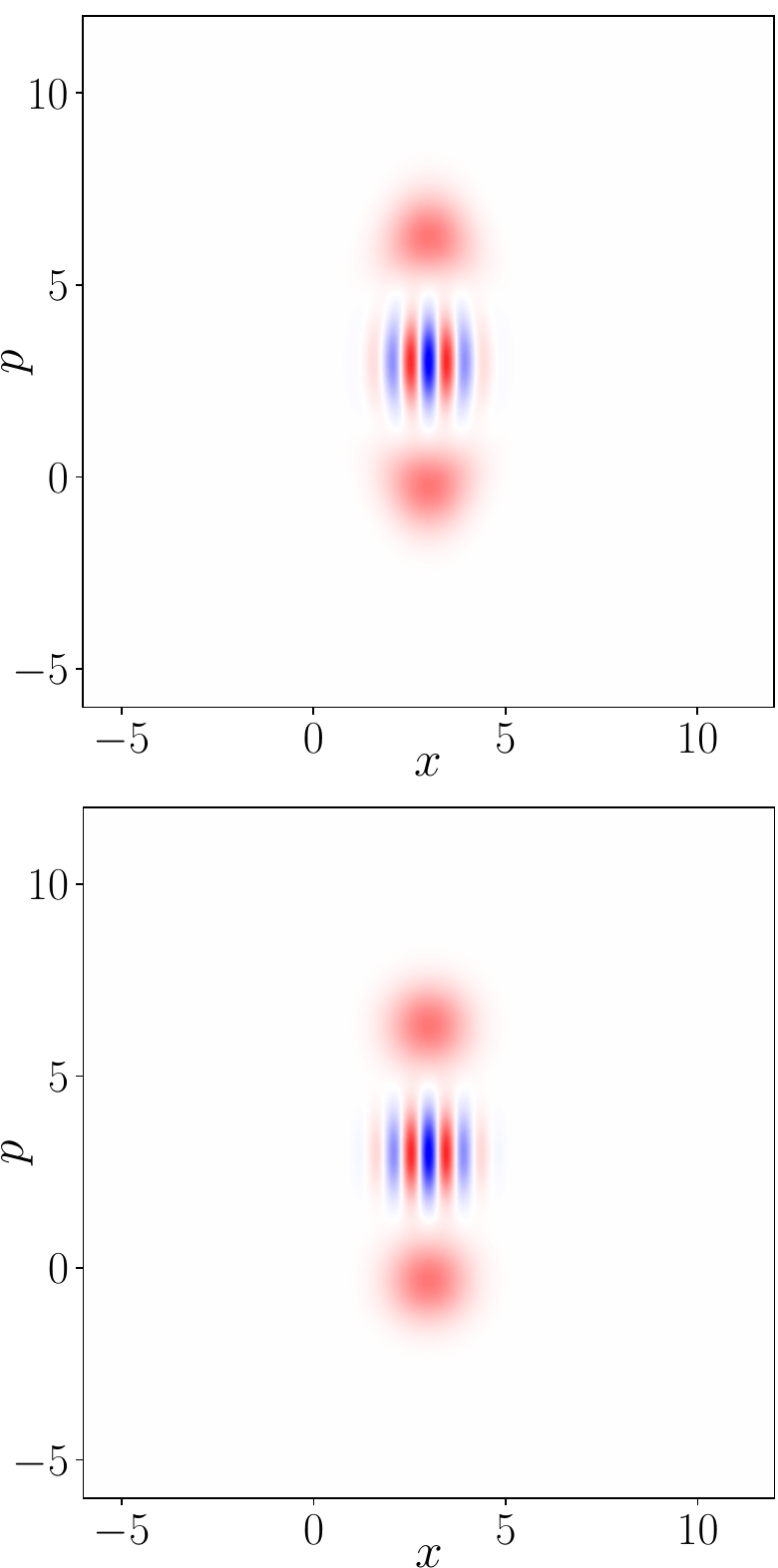} &
    \includegraphics[width=0.33189\linewidth]{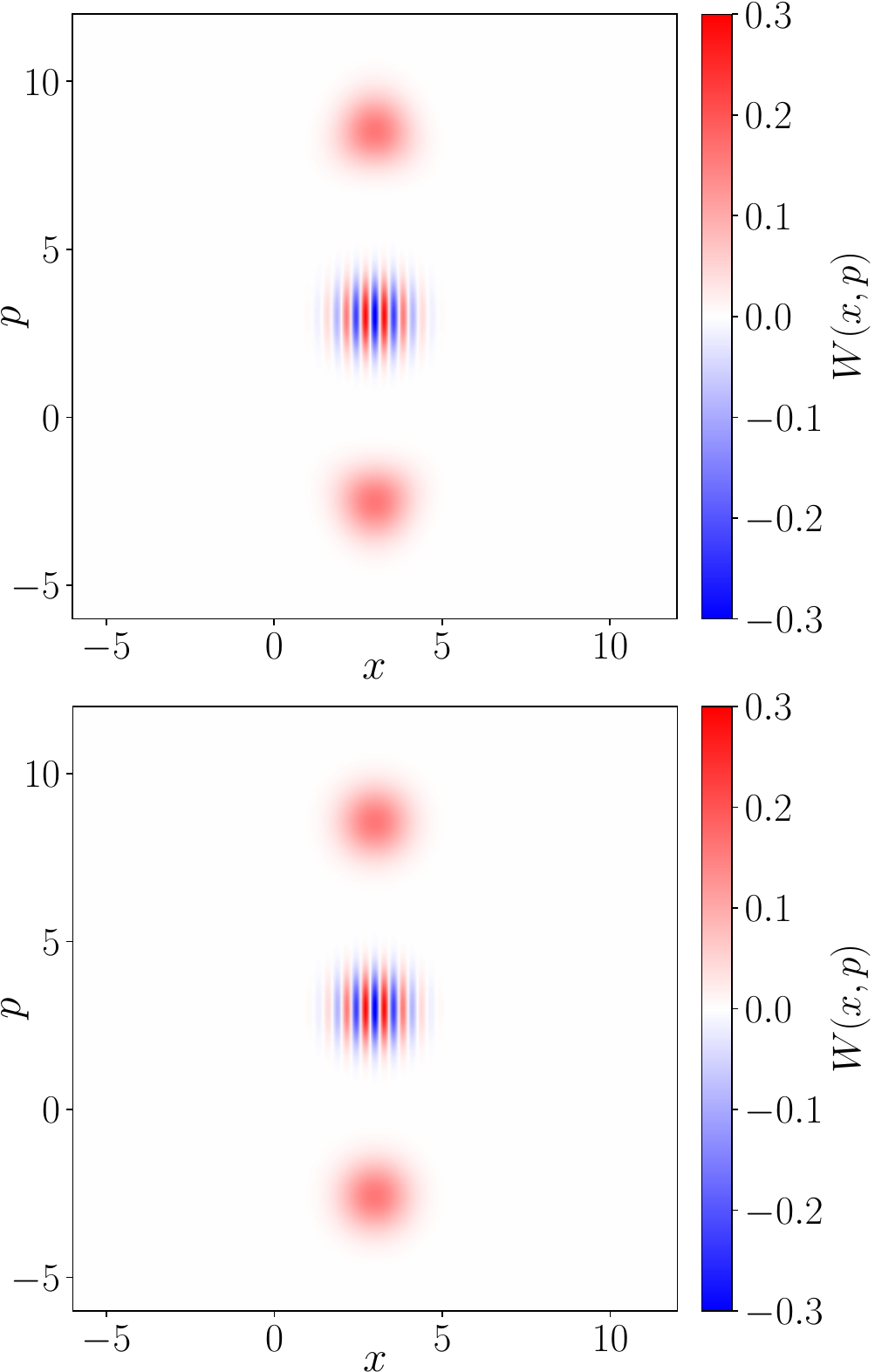}\\
    \small (a) $n=1, F_{\rm cat} =  0.9734$ & \small (b) $n=5, F_{\rm cat} =  0.9948$ & \small(c) $n=15, F_{\rm cat} =  0.9983$
  \end{tabular}
  \caption{Wigner functions of the (top row)~exact output state~\eqref{S10}  and
    (bottom row) ``perfect'' semiclassical cat~\eqref{D5}  obtained from the input coherent state with
    amplitude $\alpha_0=(3+3i)/\sqrt{2}$ corresponding to the measurement result
    $y_m=x_0=3$ for the Fock resource state with the photon number
    (a)~$n=1$; (b)~$n=5$, (c)~$n=15$. $F_{\rm cat}$ is the fidelity between the corresponding states.} 
\label{fig10}
\end{figure*}

At $\delta p^{(+)}_0=0$ when the measurement result of the
ancillary oscillator is $y_m= x_0$,
the approximate gate output state~\eqref{S101}
with the semiclassical factor
determined by formulas~\eqref{D3}--\eqref{D41}
will be
the  undistorted semiclassical Schr{\"o}dinger cat
given by
\begin{align}
  &
\psi^{\rm (out)}_{\rm cat}(x,x_0) \sim \psi^{\rm (in)}(x)[e^{i(\phi_0+p^{(+)}_0 x)}+(-1)^ne^{-i(\phi_0+p^{(+)}_0x)}],
\label{D5}
\end{align}
%(\emph{linearized approximation})
where
\begin{align}
\label{D50}
\phi_0\equiv - p^{(+)}_0 x_0,\,\,p^{(+)}_0=\sqrt{2n+1}.
\end{align}
\textcolor{black}{The} ``perfect'' Schr\"odinger cat can also be written
as a superposition of Glauber coherent states of the form:
\begin{align}
\label{D501}
|\psi^{\rm (out)}_{\rm cat}\rangle\sim  e^{i\phi_0}|\tilde{\alpha}_+\rangle + (-1)^n e^{-i\phi_0}|\tilde{\alpha}_-\rangle,
\end{align} 
where $\tilde{\alpha}_{\pm} \equiv [x_0+i(p_0\pm p^{(+)}_0)]/\sqrt{2}$.

From relations~\eqref{S10},~\eqref{S010}, and \eqref{D5}, the fidelity between
$\psi^{\rm (out)}(x,y_m)$ and $\psi^{\rm (out)}_{\rm cat}(x,x_0)$ at $y_m=x_0$
\begin{align}
  &
\label{D111}
F_{\rm cat}(n)= \left|\int dx\,\psi^{\rm (out)*}(x,x_0)\psi^{\rm (out)}_{\rm cat}(x,x_0)\right|_{y_m=x_0}^2
\end{align}
is independent of both $p_0$ and $x_0$
(since the integrand depends only on the difference
$x-x_0$ and the integral is taken over an unbounded interval)
and, thus, is determined solely by the photon number $n$.

Note that
from the semiclassical description
it can be inferred the output state created by the scheme under consideration
from the coherent state $\ket{\alpha_0}$ with the amplitude
$\alpha_0=(x_0+ip_0)/\sqrt{2}$,
at $y_m=x_0$,
will be close to the ``perfect'' cat~\eqref{D5} with
high fidelity for any photon number of the resource state
because target state is localized in proximity of $x_0$
with the characteristic length $\delta x\sim 1/\sqrt{2}$.
Therefore, for all $x$ belonging to the input function support, the semiclassical relation
$|x-x_0|\ll\sqrt{2n+1}$ will be fulfilled, which guarantees the generation of \textcolor{black}{a minimally distorted} cat state \textcolor{black}{(see Fig.~\ref{fig10}, top row)}.

\textcolor{black}{The fact that the quality of the semiclassical approximation given by
  Eq.~\eqref{D501} improves with increasing photon number $n$ is illustrated in Fig.~\ref{fig10}
  where, for $y_m=x_0=3$ and $n\in\{1,5,15\}$, the Wigner functions of the exact output
  state~\eqref{S10}) are compared with their semiclassical counterparts computed for the undeformed
  cat~\eqref{D5}.}
% The plots demonstrate high ``\textcolor{black}{proximity}'' of the gate output state~\eqref{S10} to the ``perfect''
% Schr{\"o}dinger cat~\eqref{D5},
% with fidelity $F_{\rm cat}= 0.9734,\;0.9948, \;0.9983$, respectively. 

Statistics of the homodyne measurements
of the ancilla momentum
is described by the probability density
to observe the outcome $y_m$
given by
the norm of the unnormalized output wave function~\eqref{S010}  
\begin{align}
  &
\label{DD1}
P(y_m)=\langle{\tilde{\psi}}^{\rm (out)}|{\tilde{\psi}}^{\rm (out)}\rangle=\int dx |{\tilde{\psi}}^{\rm (out)}(x,y_m)|^2,
\end{align}
so that the wave function~\eqref{S10} of the output state can be written as
\begin{align}
  &
\psi^{\rm (out)}(x,y_m) =\frac{{\tilde{\psi}}^{\rm (out)}(x,y_m)}{\sqrt{P(y_m)}}.
\label{S090}
\end{align}

\begin{figure}[t!]
\centering
\includegraphics[width=1.0\columnwidth]{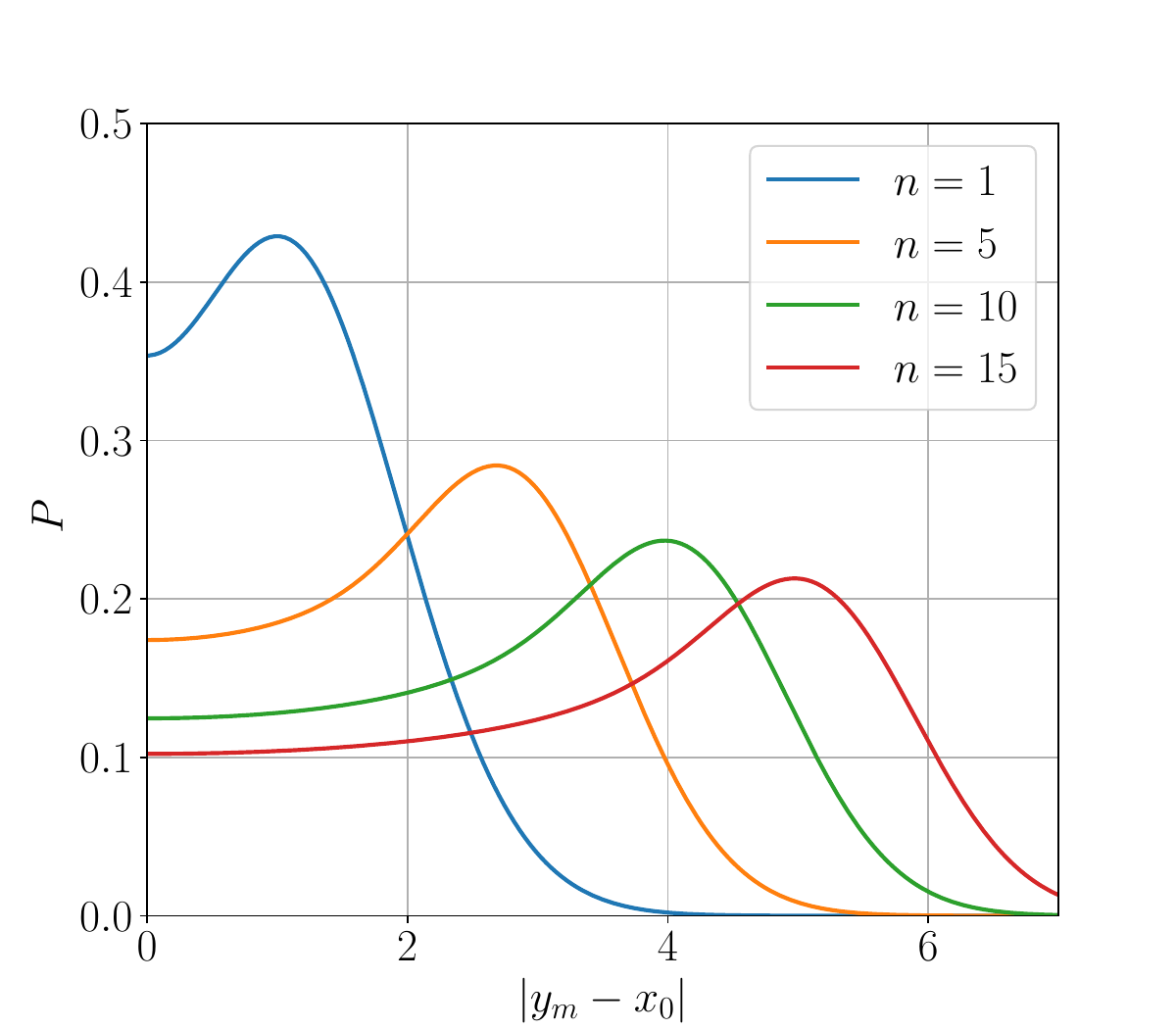}
\caption{\textcolor{black}{The probability density $P$ as a function of the displacement $|y_m - x_0|$, for different photon numbers $n \in \{1, 5, 10, 15\}$ of the Fock resource state. $P(y_m,0)$ is an even function of $y_m$ and $P(y_m,x_0)=P(y_m-x_0,0)$}} 
\label{fig099}
\end{figure}

For the input coherent state~\eqref{f65} of the target oscillator and the Fock resource
state~\eqref{S7} with $n$ photons, the probability density
takes the explicit form 
\begin{align}
  &
\label{DD01}
    P(y_m,x_0)=\frac{1}{\pi2^n n!}\int H_n^2(x-y_m) e^{-(x-y_m)^2}e^{-(x-x_0)^2} d x
    \notag
  \\
  &
    = P(y_m-x_0,0)=P(x_0-y_m,0),
\end{align}
where the probability density $P(y_m,0)$ corresponds to the vacuum state of the target
oscillator~\cite{Baeva2024}.  Note that $P(y_m,0)$ is even in $y_m$ due to the parity in $\xi$ of
the absolute value of the Hermite polynomials $|H_n(\xi)|$.  In Fig.~\ref{fig099}, the probability
density $P(y_m-x_0,0)$ is plotted as a function of $|y_m-x_0|$ at different values of the photon
number.  \textcolor{black}{Each curve shows a local maximum that decreases with $n$ while shifting outward, reflecting a fundamental quantum trade-off: higher photon numbers $n$ increase the cat-state displacement $\sqrt{2n+1}$ but broaden the ancilla's wavefunction, reducing peak success probabilities. Conversely, smaller $n$ yields higher $P$ but produces smaller cats with reduced macroscopic distinguishability.}
\textcolor{black}{The optimal regime, characterized by simultaneously moderate $P \sim 0.2\text{--}0.4$ and high fidelity $F_{\mathrm{cat}} > 0.95$, is
  achieved with small photon numbers ($n \leq 5$) and minor deviations of $y_m$ from $x_0$,
  specifically $|x_0 - y_m| \lesssim 1/\sqrt{2}$ (as illustrated for $n=1$ in Figs.~\ref{fig7}
  and~\ref{fig099}). Within this regime, the displacement offset $\Delta y_m = |y_m - x_0|$ can be
  adjusted to prioritize either $P$ or $F_{\mathrm{cat}}$, allowing for a tunable balance between
  success rate and output state quality tailored to application-specific needs.}

Since, in real-world experiments, the measurement outcome can only be identified with a certain precision,
it is instructive to consider a mixed output state that emerges
when the observed ancilla momentum falls
within an acceptance interval
ranged from $-d/2$ to $d/2$ of the width $d$
centered at $y_m=0$ provided $x_0=0$.
In this case, the weighted fidelity between the ``perfect'' semiclassical cat
state~\eqref{f9} and the mixed state at the gate output is evaluated as follows  
\begin{align}
\label{Fidelity_mix}
F^{\mathrm{mix}}(d) =\frac{1}{P^{\mathrm{mix}}(d)}\int_{-d/2}^{+d/2} dy_m\,P(y_m,0)F_{\rm cat}(y_m),
\end{align}
where
 $\displaystyle
P^{\mathrm{mix}}(d)\equiv\int_{-d/2}^{+d/2} dy_m\,P(y_m,0)
 $ 
is the probability that the measurement outcome is within the acceptance interval. 

When $x_0\ne 0$, we choose the acceptance interval to be centered at $y_m=x_0$
and insert the ideal cat~\eqref{D5} into the expression for $F_{\rm cat}(y_m)$.
After change of the variable:  $x\to x-x_0$, the expression for $F^{\mathrm{mix}}(d)$
is reduced to Eq.~\eqref{Fidelity_mix}.

\begin{figure}[t!]
\centering
\includegraphics[width=1.0\columnwidth]{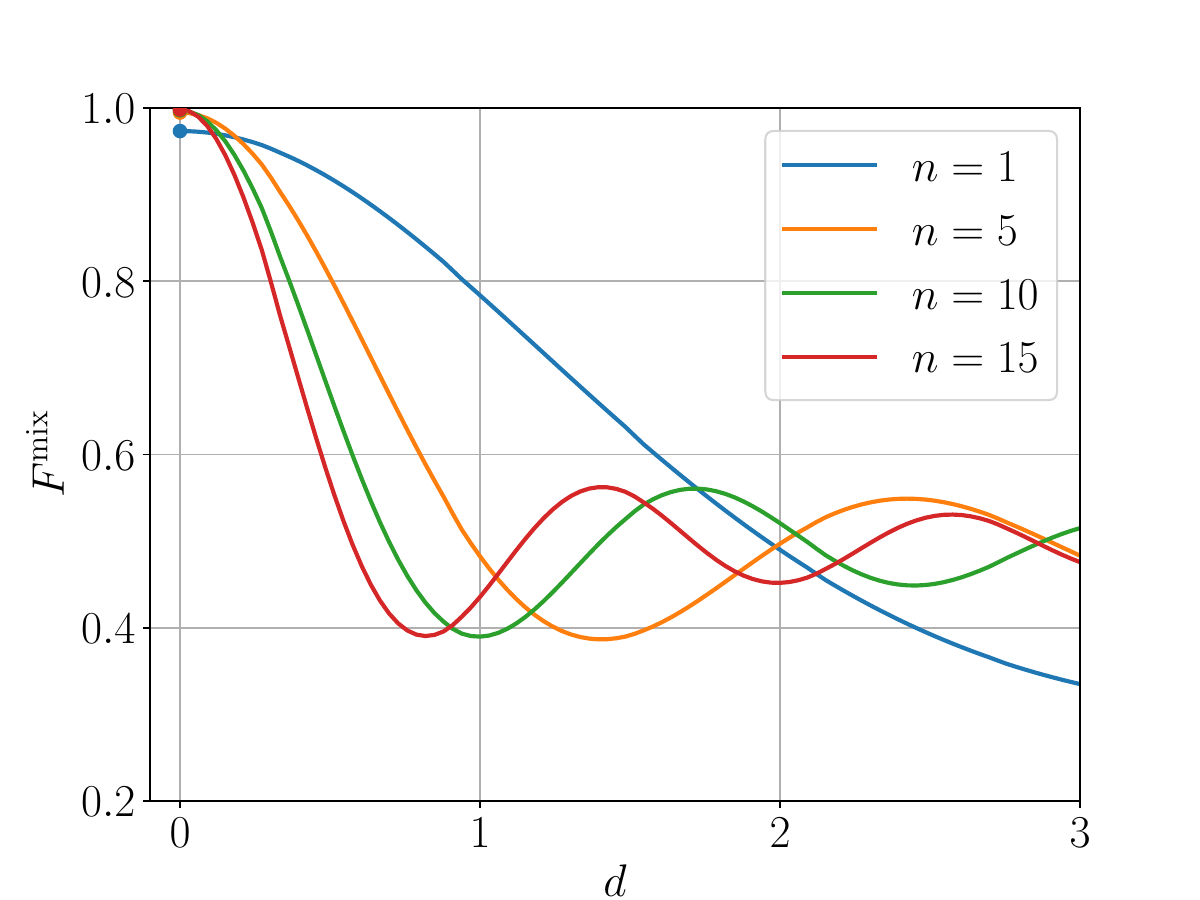}
\caption{Fidelity $F^{\mathrm{mix}}$ (see Eq.~\eqref{Fidelity_mix}) between the output mixed state and
  the ``perfect'' semiclassical cat state ~\eqref{f9} of the target oscillator as a function of the
  acceptance interval $d$ at $n\in\{1,5,10,15\}$.} 
\label{fig12}
\end{figure}

Figure~\ref{fig12} presents the results for
the fidelity $F^{\mathrm{mix}}(d)$ evaluated
as a function of the acceptance interval length, $d$,
at different values of the photon number,
$n\in \{1,5,10\}$.
From Fig.~\ref{fig12}
it is essential
to use sufficiently narrow acceptance intervals
in order to prepare a mixed state with high
fidelity $F^{\mathrm{mix}}(d)$ to the undistorted semiclassical cat state~\eqref{f9}.

It is useful to note that,
for small intervals with $d\ll\sqrt{2n+1}$,
the probability that the measurement outcome lies in the acceptance
interval is proportional to $d$
and can be estimated as $P^{\mathrm{mix}}(d)\approx P(0,0) d$.
This behavior comes from the weak dependence of the probability density
on $y_m$ at low $y_m$ (see Fig.~\ref{fig099}). 
%

%%%%%%%%%%%%%%%%%%%%%%%%%%%%%
\section{Discussions and conclusion}
\label{conclusion}

We have studied the CV measurement-induced Fock state-based ``cat gate''
which can conditionally generate a
two-component Schr{\"o}dinger cat-like superposition from the input coherent
state.
It is shown that the geometric semiclassical mapping in the phase space
can be used to evaluate such important characteristics as
the number of components of the cat-like superposition, their
positions, and distortions produced by the logical element.
We have identified the gate operation regime
in which the output state is closest to the “perfect” semiclassical Schr{\"o}dinger cat state
represented by the superposition of two undeformed copies of the input state symmetrically displaced
in the phase space.
The ``size'' of the generated Schr{\"o}dinger cat at the gate output - the ``distance''
between the copies - can be made as large as required by an appropriate choice of the gate
parameters. A key feature of the gate under consideration is that the Schr{\"o}dinger cat state
emerges when the ancillary oscillator measurement is compatible
with multiple values of the target oscillator physical variables. 

We have performed
a detailed analysis of the fidelity between the gate output state and ``perfect'' Schr{\"o}dinger
cat state derived from semiclassical theory.
We have found the criteria for the gate operation
with high fidelity values exceeding 0.99
for a coherent state at the gate input
and have illustrated the qualitative features of the output cat states
in terms of their Wigner functions depending on on the gate parameters and measurement outcome.
A clear interpretation of the output state quantum statistics in terms of the Wigner function in
dependence on the gate parameters and measurement outcome was presented. 

\textcolor{black}{For small photon numbers ($n \leq 5$) and minor deviations of the measurement outcome from the input coherent state's coordinate ($|y_m - x_0| \lesssim 1$), the protocol simultaneously achieves high fidelity $F_{\mathrm{cat}} > 0.95$ and moderate $P \sim 0.2-0.4$. This demonstrates practical viability, as multiple experimental runs can compensate for the success rate while maintaining state quality.}

In analogy to some other non-Gaussian CV schemes,
the studied ``cat gate'' uses the same key
elements such as the Fock resource state, the entangling $\hat{C}_Z$ operation, and the homodyne
measurement. A key feature of the regime where the cat-like superposition emerges is that
the projective measurement provides multivalued information about the target system's physical
variables.
From  our analysis, it might be concluded that this feature may arise in CV quantum
networks with embedded non-Gaussian gates of a general kind using more complex resource states and
other types of measurements that may cause  measurement-induced evolution leading to
the generation of the multi-component Schr{\"o}dinger cat states which can be used as the logical
qubit basis for error correction codes. The cat-breeding transformations of an arbitrary input state
may be combined with standard Gaussian operations such as displacement, rotation, squeezing, and
shear deformation, and can be successfully applied in complex non-Gaussian quantum networks. 

%%%%%%%%%%%%%%%%%%%%%%%%%%%%%
\begin{acknowledgments}

N.G.V. is grateful to Dr. Ivan V. Sokolov for the meaningful discussions. The work was supported by Russian Science Foundation (project No. 24-21-00484).

\end{acknowledgments}

\section*{Declaration of competing interests}

The authors declare that they have no known competing financial interests or personal relationships that could have appeared to influence the work reported in this paper.

\textbf{Data Availability Statement}: This manuscript has no associated data or the data will not be deposited. [Authors comment: This is a theoretical study and no experimental data has been listed.]

% \bibliography{fixed-bibliography}

\appendix
\section{Wigner Function from Mehler Formula}
\label{sec:mehler}

In this section, we describe how to calculate
the Wigner function efficiently, avoiding heavy numerical integration while maintaining high
accuracy. This approach is based on the technique of generating functions
and is especially suited for
symbolic computation environments like Maple or Mathematica
to deal with power series expansions.

We apply our method to the Wigner function~\eqref{f5} of the exact solution
given by Eqs~\eqref{S10}, \eqref{S010} and~\eqref{f65}.
To this end, the expression for the Wigner function
can be conveniently rewritten in the form
\begin{align}
  &
  \label{eq:tw}
  W_n(\tilde{x},\tilde{p})=\frac{1}{\pi \tilde{N}_n}\int_{-\infty}^{\infty} \tilde{w}_{n}(\tilde{x},z)
  e^{2 i \tilde{p} z} dz,
  \\
  &
  \tilde{w}_{n}(\tilde{x},z)=
    \frac{1}{\pi^{1/2}2^n n!}H_n(\tilde{x}-z)H_n(\tilde{x}+z)
    \notag
  \\
  &
    \times
  \exp\{
-\tilde{x}^2 - (\tilde{x}+\Delta x)^2 - 2z^2
    \},
  \\
  &
    \tilde{N}_n= \int_{-\infty}^{\infty} \tilde{w}_{n}(\tilde{x},0) d \tilde{x},
\end{align}
where $\tilde{x}=x-y_{m}$,
$\tilde{p}=p-p_{0}$ and $\Delta x=y_{m}-p_{0}$.

The well-known Mehler formula~\cite{Mehler1866,Erdelyi1953}
\begin{align}
  &
  \label{eq:Mehler}
  \sum_{n=0}^{\infty}\frac{\rho^n}{2^n n!}H_n(x_1)H_n(x_2)=
  \frac{1}{\sqrt{1-\rho^2}}
  \notag
  \\
  &
    \times
  \exp
  \bigl\{
\frac{2 \rho x_1 x_2 -\rho^2(x_1^2+x_2^2)}{1-\rho^2}
  \bigr\},
\end{align}
can now be used to evaluate the generating functions
\begin{align}
  \label{eq:mehler-norm1}
  &
    \sum_{n=0}^{\infty} \tilde{N}_{n}\rho^{n}
  =\frac{1}{\sqrt{2(1-\rho)}}e^{-\frac{\Delta x^2}{2}(1-\rho)},  
  \\
  &
      \label{eq:mehler-tw}
     \sum_{n=0}^{\infty} \rho^{n}\int_{-\infty}^{\infty} \tilde{w}_{n}(\tilde{x},z)
    e^{2 i \tilde{p} z} dz=
    \frac{e^{-\Delta x ^2/2}}{\sqrt{2}} W_0(\tilde{x},\tilde{p})
    \notag
  \\
  &
    \times
    \frac{1}{\sqrt{1+\rho}}
    \exp\Bigl\{
\frac{2\rho\tilde{x}^2}{1+\rho}+\frac{\rho\tilde{p}^2}{2}
    \Bigr\},
\end{align}
where
\begin{align}
  &
    \label{eq:W0}
    W_0(\tilde{x},\tilde{p})=\frac{1}{\pi}
    \exp\Bigl\{
-2 (\tilde{x}+\Delta x/2)^2-\frac{\tilde{p}^2}{2}
    \Bigr\}.
\end{align}
The final result reads
\begin{align}
  &
  \label{eq:mehler-Wn}
    W_n(\tilde{x},\tilde{p})=W_0(\tilde{x},\tilde{p})\frac{\tilde{W}_n(\tilde{x},\tilde{p})}{N_n},
  \\
  &
    \label{eq:mehler-tWn}
    \sum_{n=0}^{\infty}\tilde{W}_n(\tilde{x},\tilde{p}) \rho^n=
    \frac{1}{\sqrt{1+\rho}}
    \exp\Bigl\{
\frac{2\rho\tilde{x}^2}{1+\rho}+\frac{\rho\tilde{p}^2}{2}
    \Bigr\},
  \\
  &
    \label{eq:mehler-norm}
    \sum_{n=0}^{\infty} N_{n}\rho^{n}
  =\frac{e^{\frac{\rho \Delta x^2}{2}}}{\sqrt{1-\rho}}.
\end{align}

For the photon number $n$,
formula~\eqref{eq:mehler-Wn} gives the Wigner function $W_n$
expressed terms of the function $\tilde{W}_n$ and the normalization coefficient $N_n$
that can be found as coefficients of power series expansions in $\rho$ of the corresponding
generating functions. The right-hand sides
of Eqs.~\eqref{eq:mehler-tWn} and~\eqref{eq:mehler-norm}
provide analytical expressions for these functions.

% When calculating the Wigner function~\eqref{f5} directly, a similar procedure can be performed. The unnormalized contribution is given by:
% % \begin{widetext}

% Normalizing  to  found in Eq.~\eqref{eq:mehler-norm} yields the exact result:

% Thus, to find the Wigner function for the required number of photons $n$, it is sufficient to
% determine the corresponding term in the series expansion.
\end{document}